\documentclass[preprint,showpacs,preprintnumbers,amsmath,amssymb]{revtex4}
\usepackage{graphicx,epsfig,latexsym,makeidx,amsmath}
\usepackage{color}
\usepackage{float}
\usepackage{subcaption} 
\usepackage{amssymb} 
\usepackage{caption}    
\usepackage[colorlinks,linkcolor=blue,anchorcolor=blue,citecolor=green,urlcolor=blue]{hyperref}
\usepackage[english]{babel}
\usepackage{booktabs}  
\usepackage{array}     
\usepackage{diagbox}    
\usepackage{multirow}

\usepackage[a4paper,top=2cm,bottom=2cm,left=3cm,right=3cm,marginparwidth=1.75cm]{geometry}
\begin{document}
	\newcommand \nn{\nonumber}
	\newcommand \fc{\frac}
	\newcommand \lt{\left}
	\newcommand \rt{\right}
	\newcommand \pd{\partial}
	\newcommand \e{\text{e}}
	\newcommand \hmn{h_{\mu\nu}}
	\newcommand{\PR}[1]{\ensuremath{\left[#1\right]}} 
	\newcommand{\PC}[1]{\ensuremath{\left(#1\right)}} 
	\newcommand{\PX}[1]{\ensuremath{\left\lbrace#1\right\rbrace}} 
	\newcommand{\BR}[1]{\ensuremath{\left\langle#1\right\vert}} 
	\newcommand{\KT}[1]{\ensuremath{\left\vert#1\right\rangle}} 
	\newcommand{\MD}[1]{\ensuremath{\left\vert#1\right\vert}} 

	\title{Quasinormal modes of thick branes in $f(R,T)$ gravity}

	\author{Yu-Peng E$^{a}$ $^{b}$, Hai-Long Jia$^{a}$ $^{b}$, and Chun-Chun Zhu$^{c}$ \footnote{202507008@jcut.edu.cn, corresponding author}}
	
	\affiliation{$^{a}$Lanzhou Center for Theoretical Physics, Key Laboratory for Quantum Theory and Applications of the MoE, Key Laboratory of Theoretical Physics of Gansu Province,  Gansu 
		Provincial Research Center for Basic Disciplines of Quantum Physics, Lanzhou University, Lanzhou 730000, China\\
		$^{b}$Institute of Theoretical Physics and Research Center of Gravitation, School of Physical Science and Technology,  Lanzhou University, Lanzhou 730000, China\\
		$^{c}$School of Mathematics and Physics, Jingchu University of Technology, Jingmen 448000, China\\}
	\begin{abstract} 
We study tensor perturbations and quasinormal modes (QNMs) of thick branes in $f(R,T)=R+\alpha T$ gravity. Starting from the background thick brane solutions, we derive the Schr\"odinger-like equation for gravitational perturbations and construct the corresponding effective potential. The effective potential exhibits a double-barrier structure whose properties depend on the model parameters. We analyze transmission spectra and find good agreement between resonance peaks and the real parts of the quasinormal frequencies. The parameters $b$ and $\alpha$ mainly control the width and height of the potential barriers, respectively. QNMs are extracted using both time-domain evolution and direct integration methods, showing consistent results for low-lying modes. Higher overtones become increasingly difficult to extract due to stronger damping and numerical contamination. Our results indicate that $f(R,T)$ corrections significantly affect both the effective potential and the quasinormal spectrum, providing a possible imprint of modified gravity in braneworld perturbations.
	\end{abstract}
	
	\maketitle
	\flushbottom
	
\section{Introduction}
    \label{sec:introduction}
The late-time accelerated expansion of the universe and the anomalous rotation curves of galaxies pose two important challenges to the standard theory of gravity \cite{Riess1998,Perlmutter1999,Peebles2003,deBernardis2000,Hanany2000,Padmanabhan2003}. These observational results strongly suggest that, on cosmological scales, either an unknown matter component must be introduced or general relativity needs to be modified~\cite{NOJIRI_2007,BransDicke1961,Horndeski1974,Clifton2012ModifiedGravity,De_Felice_2010}. In this context, a large class of modified gravity theories has been proposed to provide a unified framework for explaining these phenomena.

Among these theories, $f(R)$ gravity is one of the simplest extensions of Einstein’s theory, in which the Ricci scalar $R$ in the gravitational action is generalized to an arbitrary function $f(R)$ \cite{Carroll2004,Sotiriou2010,De_Felice_2010}. This generalization can naturally give rise to the late-time accelerated expansion of the universe without explicitly introducing dark energy. More general frameworks include non-minimal couplings between geometry and matter fields, providing richer physical structures and allowing for a more flexible description of gravitational interactions \cite{Sotiriou2010,Nojiri2011,Carroll_2005,Harko2011,Haghani:2013oma,Alvarenga_2013}.

On the other hand, higher-dimensional theories provide an alternative perspective on fundamental interactions. The idea of extra spatial dimensions can be traced back to early attempts to unify gravity and electromagnetism, in particular the early work of Nordström, and was later systematically developed in the Kaluza–Klein (KK) theory by Kaluza and Klein \cite{Kaluza1921,Klein1926}. These ideas opened a new direction in higher-dimensional physics, suggesting that gauge interactions may originate from higher-dimensional geometric structures.

In this context, braneworld models provide an important realization of extra-dimensional physics and offer a possible approach to addressing the hierarchy problem between the Planck scale and the electroweak scale. The Randall–Sundrum (RS) models, particularly RS-I and RS-II \cite{5,Randall:1999vf}, are among the most influential constructions in this direction. In the RS-I model, the extra dimension is compactified between two branes embedded in a five-dimensional anti-de Sitter spacetime. In the RS-II model, one of the branes is pushed to infinity, resulting in a single-brane configuration with a non-compact extra dimension. Notably, despite the presence of an infinite extra dimension, four-dimensional Newtonian gravity can still be recovered on the brane due to the warped spacetime geometry.

However, both the RS-I and RS-II models are based on the idealized assumption of zero-thickness branes. To describe the internal structure of the brane more realistically, thick brane models coupled to scalar fields have been introduced, leading to smooth brane configurations \cite{6,7,8,9}. Such constructions can be interpreted as domain-wall-like solutions in higher-dimensional spacetime and have been extensively studied in various modified gravity theories and in studies of matter-field localization \cite{Afonso:2007gc,Deruelle:2007pt,Borzou:2009gn,Dzhunushaliev:2009dt,Zhong:2010ae,HoffdaSilva:2011si,Balcerzak:2011bx,Carames:2012gr,Bazeia:2013oha,Bazeia:2013uva,Chakraborty:2014xla,Bazeia:2014poa,Gu:2014ssa,Zhong:2015pta,Chen:2017diy,Cui:2018hqx,Gu:2018lub,Dzhunushaliev:2019wvv,Dzhunushaliev:2019ysu,Guo:2019vvm,Cui:2020fiz}. In particular, thick brane models within the framework of $f(R,T)$ gravity have also been investigated in the literature \cite{Gu2017,Moraes2015,Rosa2021,Rosa2021Cuscuton}.

Beyond background solutions, the stability and dynamical behavior of perturbations are also crucial for understanding such systems. In black hole physics, perturbations give rise to characteristic damped oscillations known as quasinormal modes (QNMs), whose corresponding complex frequencies are referred to as quasinormal frequencies (QNFs) \cite{Vishveshwara1970,Press1971,Davis1971}. These modes carry important information about the spacetime geometry and its stability, and have been extensively studied in various gravitational systems.

Interestingly, similar quasinormal behavior also appears in braneworld models. When the brane is perturbed, fluctuations of the gravitational field or matter fields may also exhibit damped oscillatory behavior analogous to QNMs \cite{Zhong:2017uhn,Guo:2018tpo}. This raises the question of whether such quasinormal modes also exist in braneworld scenarios constructed within $f(R,T)$ gravity and how the coupling between matter and geometry affects their properties. Motivated by this question, we investigate thick brane solutions in the framework of $f(R,T)=R+\alpha T$ gravity and study their associated QNMs.

The organization of this paper is as follows. In Sec.~\ref{sec:method}, we review the thick brane background in $f(R,T)$ gravity and derive the tensor perturbation equation. In Sec.~\ref{Background solutions and  quasinormal modes}, we compute the quasinormal frequencies using both the numerical evolution method and the direct integration method, and systematically analyze the influence of the parameter $\alpha$ on the QNMs. Finally, our conclusions and discussion are presented in Sec.~\ref{sec:conclusion}.

\section{ Background and Linear Perturbations }
    \label{sec:method}
In this section, we briefly review five-dimensional thick brane models in $f(R,T)$ gravity with a minimally coupled scalar field. The action is given by
\begin{equation}
    S = \int d^5x \sqrt{-g} \left( \frac{1}{2\kappa_5^2}f(R,T)+L_m\right),
\end{equation}
where $L_m = - \frac{1}{2} g^{MN} \nabla_M \phi \nabla_N \phi - V(\phi)$, $R$ is the Ricci scalar, and $T = g^{MN} T_{MN}$ denotes the trace of the energy-momentum tensor. The stress-energy tensor is defined as
\begin{equation}
    T_{MN} = -\frac{2}{\sqrt{-g}} \frac{\delta (\sqrt{-g} L_m)}{\delta g^{MN}}.
\end{equation}

By varying the action with respect to the metric $g_{MN}$ and the scalar field $\phi$, one obtains the gravitational field equations
\begin{equation}
\begin{aligned}
f_{R} R_{MN} + \left( g_{MN} \Box^{(5)} - \nabla_M \nabla_N \right) f_{R}
- \frac{1}{2} g_{MN} f
= -f_{T} \left( T_{MN} + \Theta_{MN} \right) + \kappa_5^2 T_{MN},\label{G}
\end{aligned}
\end{equation}
and the scalar field equation
\begin{equation}
3 \nabla^M \left( f_T \nabla_M \phi \right) + 2\kappa_5^2 \Box^{(5)} \phi 
= \left( 5f_T + 2\kappa_5^2 \right) V_{\phi},\label{S}
\end{equation}
where $f_R=\partial f/\partial R$, $f_T=\partial f/\partial T$, and $V_{\phi}=dV/d\phi$. The tensor $\Theta_{MN}$ is defined as
\begin{equation}
 \Theta_{MN} = g^{AB}\frac{\delta T_{AB}}{\delta g^{MN}}.
\end{equation}
In the following, we set $\kappa_5 = 1$ for simplicity.

We consider a five-dimensional warped metric of the form
\begin{equation}
    ds^2 = e^{2A(y)} \eta_{\mu\nu} dx^\mu dx^\nu + dy^2,
    \label{metric}
\end{equation}
where $e^{A(y)}$ is the warp factor $a(y)$. Introducing the conformal coordinate $z$ via $dy=e^{A(z)}dz$, the metric becomes
\begin{equation}
    ds^2 = e^{2A(z)}\left( \eta_{\mu\nu} dx^\mu dx^\nu + dz^2 \right).
\end{equation}

Substituting the metric \eqref{metric} into the field equations \eqref{G} and~\eqref{S} yields the background equations
\begin{equation}
    f_{R}(4A'^2+A'')- f''_{R} - 3A'f'_{R}+\frac{1}{2}f
    = \frac{1}{2}\phi'^2 + V,
\end{equation}
\begin{equation}
-4f_{R}(A'^2+A'')+4 A' f'_R-\frac{1}{2}f
 = \frac{3}{2} f_{T} \phi'^2 + \frac{1}{2}\phi'^2 - V,
\end{equation}
\begin{equation}
  3f'_T\phi' + (3f_T+2)(4A^{\prime}\phi^{\prime}+\phi^{\prime \prime})
  = (5f_T+2)V_\phi,
\end{equation}
Here and in the background equations below, a prime denotes differentiation with respect to $y$. Derivatives with respect to the conformal coordinate $z$ are written explicitly using $\partial_z$.

Next, we consider tensor perturbations of the metric in the coordinates $(x^{\mu},z)$:
	\begin{equation}
		\displaystyle
		ds^2 = e^{2A(z)} \left( (\eta_{\mu\nu} + h_{\mu\nu}) dx^\mu dx^\nu + dz^2 \right),
		\label{9}
	\end{equation}
	where $ h_{\mu\nu} $ is a transverse-traceless tensor:
	\begin{equation}
		\displaystyle
	     \partial^\mu h_{\mu\nu} = 0, 
		~\eta^{\mu\nu} h_{\mu\nu} = 0.
		\label{10}
	\end{equation}
	The tensor perturbation can be decomposed as follows:
	\begin{equation}
		h_{\mu\nu}(x^i,t,z) = \left( a^{-\frac{3}{2}} f_R^{-\frac{1}{2}} \right) \epsilon_{\mu\nu}(x^i) \psi(t,z),
		\label{11}
	\end{equation}
	where $\delta^{ij} \partial_i \partial_j \epsilon_{\mu\nu} = -p^2 \epsilon_{\mu\nu}$ with $p^2=\delta_{ij} p^i p^j$ denoting the square of the three-dimensional momentum of the gravitational Kaluza-Klein mode. By substituting Eqs.~\eqref{9} and \eqref{11} into Eq.~\eqref{G}, we obtain the evolution equation for the extra-dimensional component:
	\begin{equation}
		\partial_t^2 \psi(t, z) - \partial_z^2 \psi(t, z) +W(z) \psi(t, z) = -p^2 \psi(t, z).
        \label{waveeq}
	\end{equation}
	Here, the effective potential $W(z)$ is~\cite{38}
	\begin{equation}
		\displaystyle
		W(z) = \frac{3}{2}\partial_z^2 A(z)+\frac{9}{4}(\partial_z A(z))^2.
        \label{effp}
	\end{equation}
	Note that no explicit scalar-field terms appear in Eqs.~\eqref{waveeq} and~\eqref{effp}. The reason is that, within the framework considered in this paper, the scalar field is minimally coupled to gravity, and we focus only on transverse-traceless tensor perturbations. More specifically, the scalar field affects the system only through the background equations, and its dynamical degrees of freedom do not directly couple to the TT tensor perturbations. As a result, when deriving the evolution equation for the extra-dimensional component from Eq.~\eqref{waveeq}, the scalar field does not appear as an explicit source term.
    
    Assuming $\psi(t, z) = \varphi(z) e^{-i \omega t}$, this equation reduces to the Schr\"odinger-like form \cite{38}:
	\begin{equation}
		\left(-\partial^2_z+ W(z)\right)\varphi(z) = m^2 \varphi(z),
		\label{13}
	\end{equation}
	where $ m^2 = \omega^2 - p^2$ is the four-dimensional effective mass squared of the Kaluza-Klein mode. 
	The Schrödinger-like equation \eqref{13} can be rewritten as
	\begin{equation}
		\displaystyle
		\mathcal{K} \mathcal{K}^\dagger \varphi(z) = m^2 \varphi(z),
	\end{equation}
	where the two operators are given by
	\begin{align}
		\mathcal{K} & =\partial_z + \frac{3}{2} \frac{\partial_z a}{a}, \\
		\mathcal{K}^\dagger & =-\partial_z + \frac{3}{2} \frac{\partial_z a}{a}.
	\end{align}
	By setting $m=0$, we obtain the gravitational zero mode:
	\begin{equation}
		\displaystyle
		\varphi^{(0)}(z) = N_0 a^{3/2}(z).
	\end{equation}
	It is easy to show that the gravitational zero mode for the $f(R,T)$-brane found in Ref.~\cite{36}  satisfies
	\begin{equation}
		\displaystyle
		\int_{-\infty}^{\infty} |\varphi^{(0)}(z)|^2 \, dz < \infty.
	\end{equation}
	This result indicates that the zero mode is normalizable and hence is localized on the brane. In the next section, we consider two background solutions and calculate the corresponding QNMs under tensor perturbations.

\section{Background solutions and  quasinormal modes}
\label{Background solutions and  quasinormal modes}
In this section, we consider the following $f(R,T)$ model,
\begin{equation}
f(R,T)=R+\alpha T,
\label{f1}
\end{equation}
where $\alpha$ is a constant parameter. The Ricci scalar and the trace of the energy-momentum tensor are given by
\[
R=-4\left(5A^{\prime2}+2A^{\prime\prime}\right),\quad
T=-\frac32\phi^{\prime2}-5V.
\]
The corresponding background field equations are
\begin{equation}
-6A^{\prime2}-3A^{\prime\prime}
=\frac{2 +3\alpha}{4}\phi^{\prime2}+\frac{2 +5\alpha}{2}V,
\end{equation}
\begin{equation}
6A^{\prime2}
=
\frac{2 +3\alpha}{4}\phi^{\prime2}
-
\frac{2 +5\alpha}{2}V,
\end{equation}
\begin{equation}
(3\alpha+2 )
\left(4A^\prime\phi^\prime+\phi^{\prime\prime}\right)=(5\alpha+2 )V_\phi.
\end{equation}
By combining the above field equations, we obtain
\begin{equation}
\phi'^2(y)=-\frac{6}{2 +3\alpha}A''(y).
\end{equation}
\begin{equation}
V(y)=-\frac{12}{2 +5\alpha}A'^2(y)-\frac{3}{2 +5\alpha}A''(y),
\label{1Vy}
\end{equation}
Since the scalar field considered here is real, we require $\phi'^2(y)>0$, which leads to 
\begin{equation}
\alpha>-\frac{2 }{3}.
\label{a1}
\end{equation}
\subsection{Solution I}
\begin{figure}[htbp]  
    \centering        
    \includegraphics[width=0.5\textwidth]{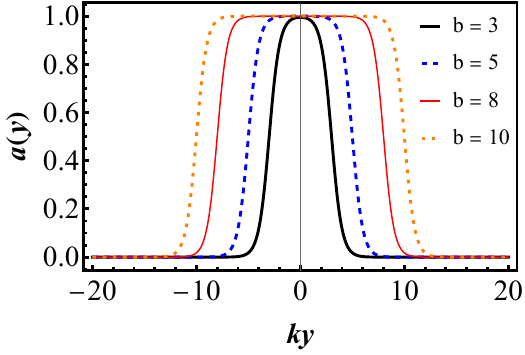}
    \caption{ The warp factor of Solution I with different values of the parameter $b$.}
    \label{pingtai a}  
\end{figure}

\begin{figure}[ht]
		\centering
		\begin{subfigure}{0.45\textwidth} 
			\centering
			\includegraphics[width=\linewidth]{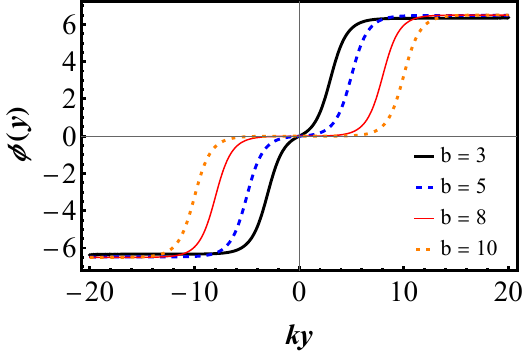} 
			\caption{$\alpha=-0.2$}
			\label{2(b)}
		\end{subfigure}
		\hspace{0.025\textwidth} 
		\begin{subfigure}{0.45\textwidth} 
			\centering
			\includegraphics[width=\linewidth]{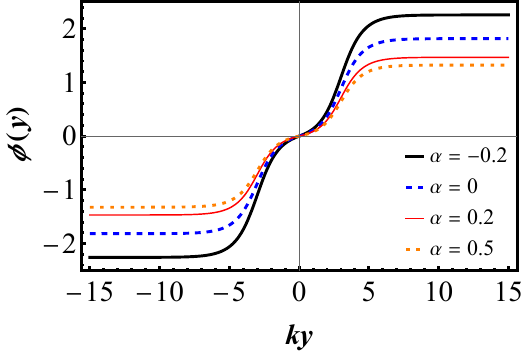} 
			\caption{$b=3$}
			\label{2(c)}
		\end{subfigure}
		\vspace{0.5cm} 
		\begin{subfigure}{0.45\textwidth} 
			\centering
			\includegraphics[width=\linewidth]{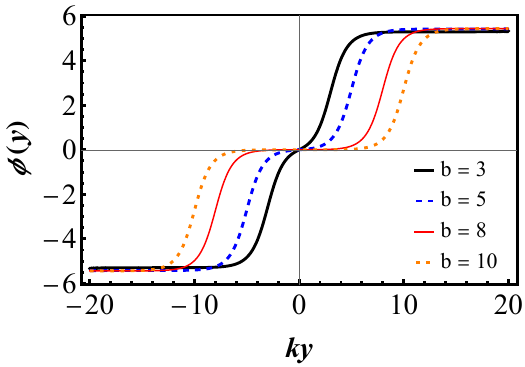} 
			\caption{$\alpha=0$}
			\label{2(d)}
		\end{subfigure}
		\hspace{0.025\textwidth} 
		\begin{subfigure}{0.45\textwidth} 
			\centering
			\includegraphics[width=\linewidth]{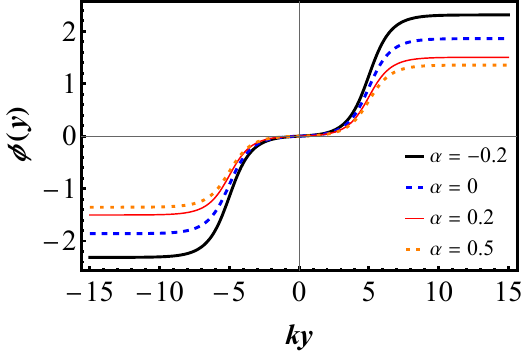} 
			\caption{$b=5$}
			\label{2(e)}
		\end{subfigure}
		\vspace{0.5cm} 
		\begin{subfigure}{0.45\textwidth} 
			\centering
			\includegraphics[width=\linewidth]{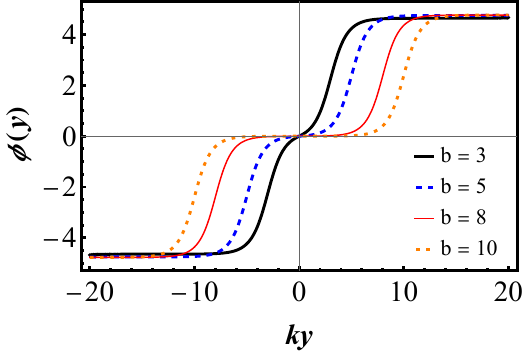} 
			\caption{$\alpha=0.2$}
			\label{2(f)}
		\end{subfigure}
		\hspace{0.025\textwidth} 
		\begin{subfigure}{0.45\textwidth} 
			\centering
			\includegraphics[width=\linewidth]{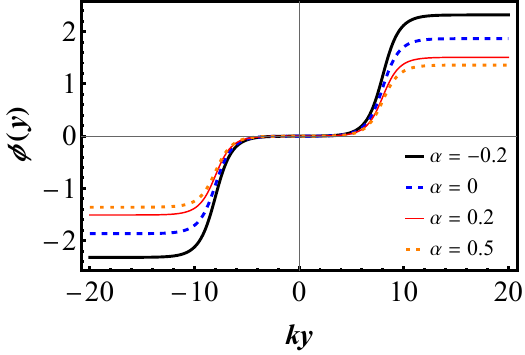} 
			\caption{$b=8$}
			\label{2(g)}
		\end{subfigure}
		\caption{The scalar field $\phi(y)$ in Solution I with different values of the parameters $b$ and $\alpha$.
		}
		\label{modelAsclar}
	\end{figure}

A plateau-like solution can be written as
\begin{align}
A(y) &= \log\left( \tanh(ky + b) - \tanh(ky - b) \right) - \log 2,\label{1Ay}\\
\phi(y)&=
\frac{2\sqrt6}{\sqrt{-(2+3\alpha)}}
\sqrt{\frac{Y(y)}{1+\cosh(2bk)}}
\left(\cosh(2bk)F(y)-2\sinh^2(bk)\Pi(y))\right),\\
Y(y)&=1+\cosh(2bk)\cosh(2ky),\notag\\
F(y)&=\int_0^{iky}\frac{d\theta}{\sqrt{1-\left[1+\tanh^2(bk)\right]\sin^2\theta}},\notag\\
\Pi(y)
&=
\int_0^{iky}
\frac{d\theta}
{\left[1-\operatorname{sech}^2(bk)\sin^2\theta\right]
\sqrt{1-\left[1+\tanh^2(bk)\right]\sin^2\theta}},
\notag \\
V(y)&=
\frac{3k^{2}\operatorname{sech}^{2}\!\left(k(y-b)\right)
\operatorname{sech}^{2}\!\left(k(y+b)\right)}{4+10\alpha}
\\
&\quad \times\left(
6 - 4\cosh(4ky)
+ \cosh\!\left(2k(y-b)\right)
+ \cosh\!\left(2k(y+b)\right)
\right).
\end{align}
where $b$ is a parameter that determines the width of the plateau, as shown in Fig.~\ref{pingtai a}. 
Taking into account the $ \mathbb{Z}_2 $ symmetry of the background, we choose a kink configuration for the scalar field satisfying $ \phi(0)=0 $. In this configuration, the scalar field exhibits a double-kink profile. The parameter $b$ also represents the distance between the two sub-kinks, and the vacuum expectation value of the scalar field decreases with the parameter $\alpha$, as shown in Fig.~\ref{modelAsclar}. 

\begin{figure}[htbp]
    \centering
        \includegraphics[width=0.5\textwidth]{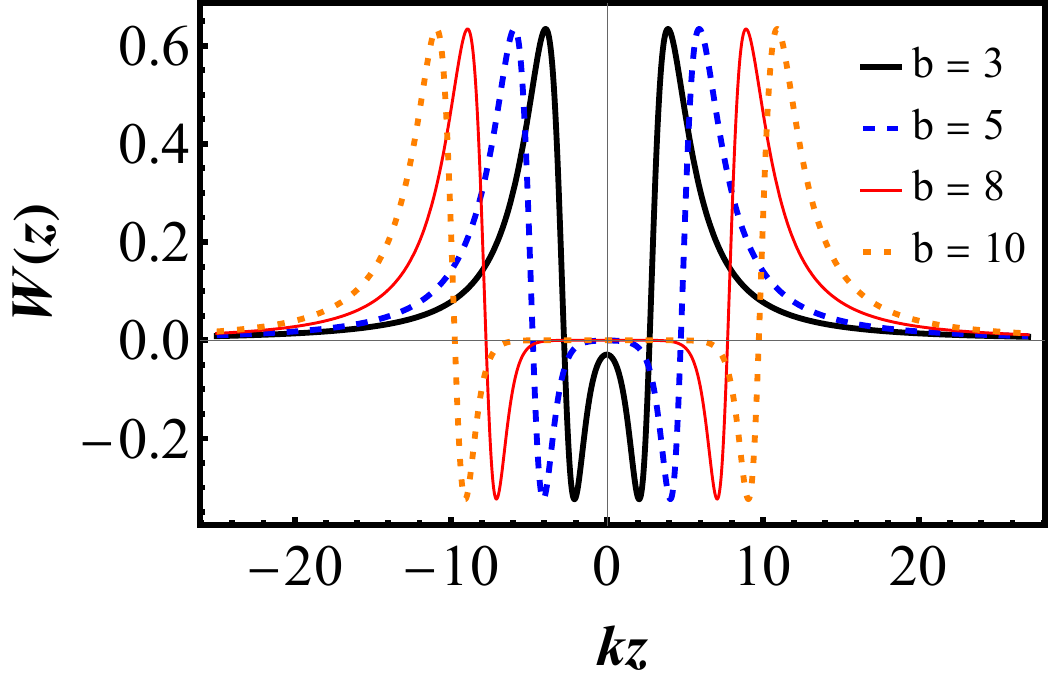} 
        \label{7a}
    \caption{The shape of the effective potential $W(z)$ for different values of $b$.}
    \label{n1平台有效势}  
\end{figure}
 
\begin{figure}[htbp]
    \centering
    \hspace{0.0125\textwidth}
    \begin{subfigure}[b]{0.45\textwidth}
        \centering
        \includegraphics[width=\textwidth]{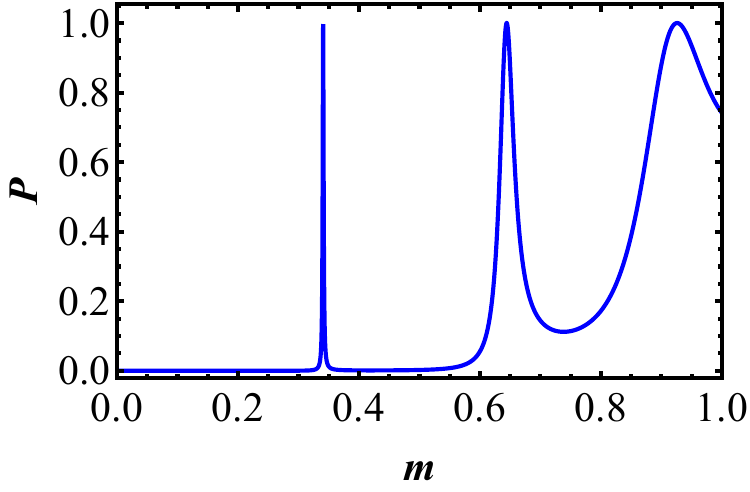} 
            \caption{$b=5$}
    \end{subfigure}
        \hspace{0.0125\textwidth} 
    \begin{subfigure}[b]{0.45\textwidth}
        \centering
        \includegraphics[width=\textwidth]{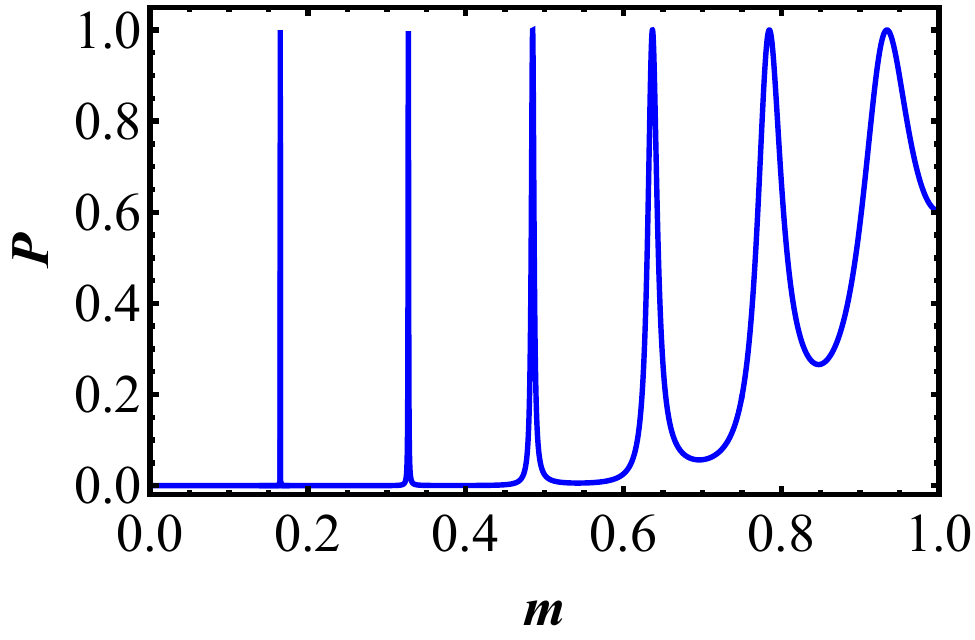} 
        \caption{$b=10$}
    \end{subfigure}
    \caption{The transmission spectrum of the effective potential in Solution I.}
    \label{对比图}  
\end{figure}

The effective potential can be obtained from the background solution, as shown in Fig.~\ref{n1平台有效势}. The effective potential for tensor perturbations exhibits a typical double-barrier structure. The transmission spectrum of such a potential can be computed using the transfer matrix method. Since the peaks of the transmission spectrum are closely related to the real parts of the QNFs~\cite{Tan:2023cra,Seahra:2005wk,E:2025kic}, the mass $m$ corresponding to these peaks can be used to construct the associated wave function as an initial profile for time evolution. This procedure effectively suppresses contributions from unwanted overtones. Consequently, the oscillation frequency and decay rate of the QNMs can be extracted.

\begin{figure}[htbp]
		\centering
		\begin{subfigure}{0.44\textwidth} 
			\centering
			\includegraphics[width=\linewidth]{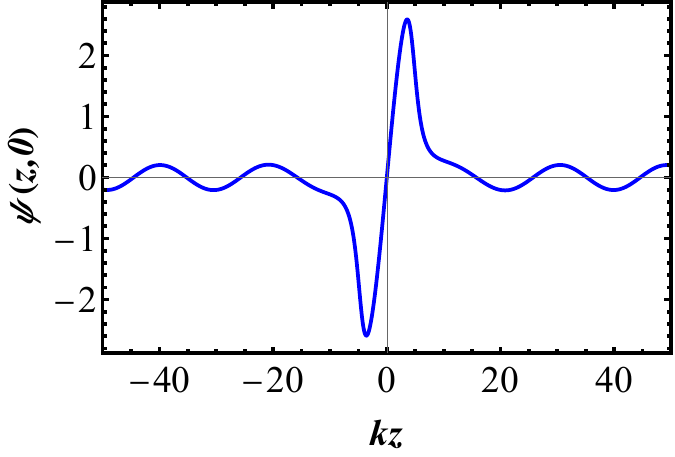} 
			\caption{Initial wave profile, $n=1$}
			\label{5(b)}
		\end{subfigure}
		\hspace{0.025\textwidth} 
		\begin{subfigure}{0.45\textwidth} 
			\centering
			\includegraphics[width=\linewidth]{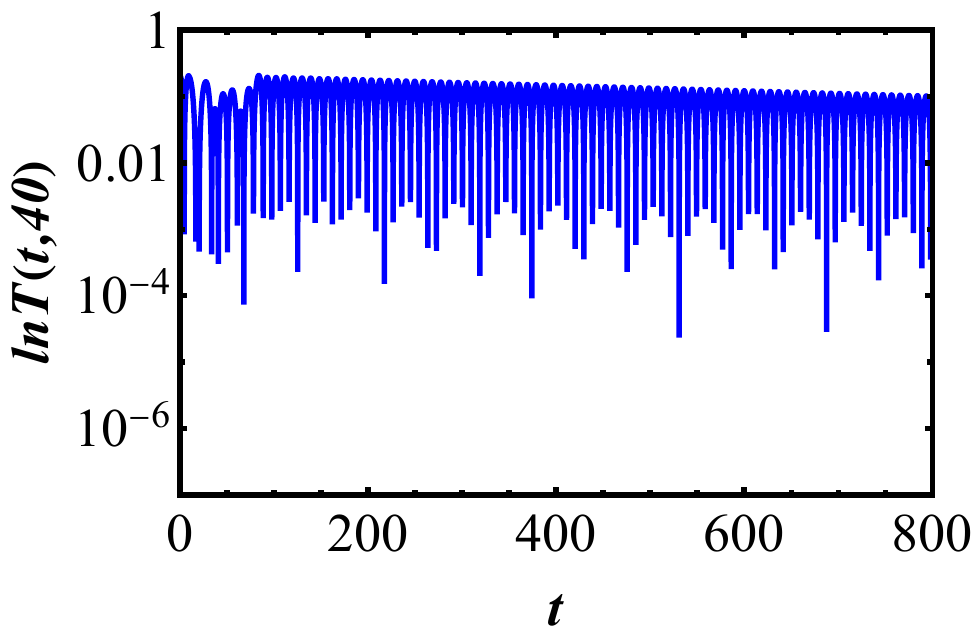} 
			\caption{ $T(t,40) = |\psi (t,40)|,n=1$}
			\label{5(c)}
		\end{subfigure}
		\vspace{0.5cm} 
		\begin{subfigure}{0.46\textwidth} 
			\centering
			\includegraphics[width=\linewidth]{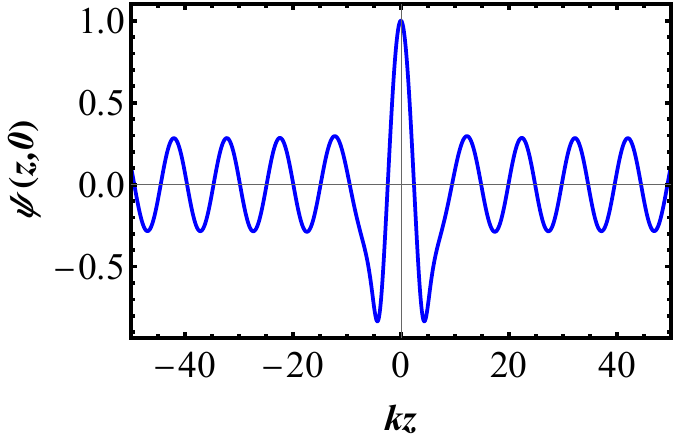} 
			\caption{Initial wave profile, $n=2$}
			\label{5(d)}
		\end{subfigure}
		\hspace{0.025\textwidth} 
		\begin{subfigure}{0.45\textwidth} 
			\centering
			\includegraphics[width=\linewidth]{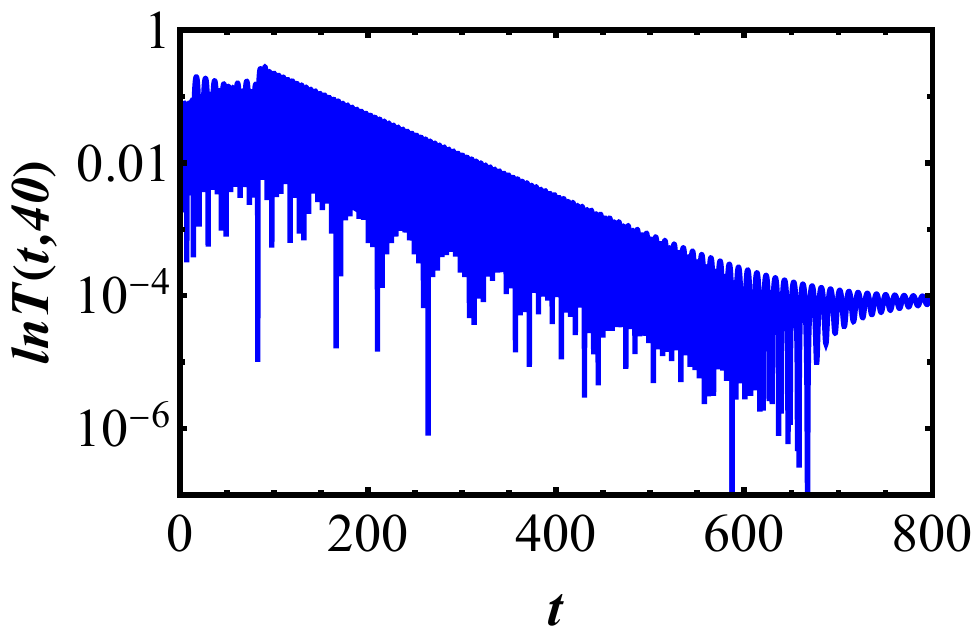} 
			\caption{$T(t,40) = |\psi (t,40)|,n=2$}
			\label{5(e)}
		\end{subfigure}
		\caption{The left column shows the initial wave profile, while the right column presents the logarithmic time evolution of the waveform.}
		\label{波形演化}
	\end{figure}

\begin{figure}[htbp]
		\centering
		\begin{subfigure}{0.45\textwidth} 
			\centering
			\includegraphics[width=\linewidth]{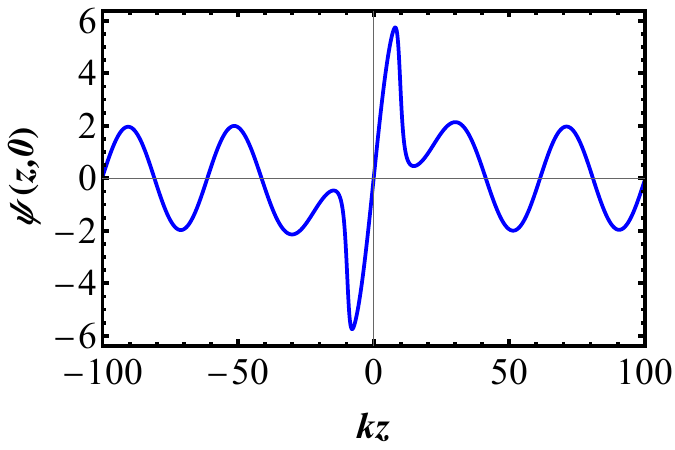} 
			\caption{Initial wave profile, $n=1$}
			\label{6(b)}
		\end{subfigure}
		\hspace{0.025\textwidth} 
		\begin{subfigure}{0.45\textwidth} 
			\centering
			\includegraphics[width=\linewidth]{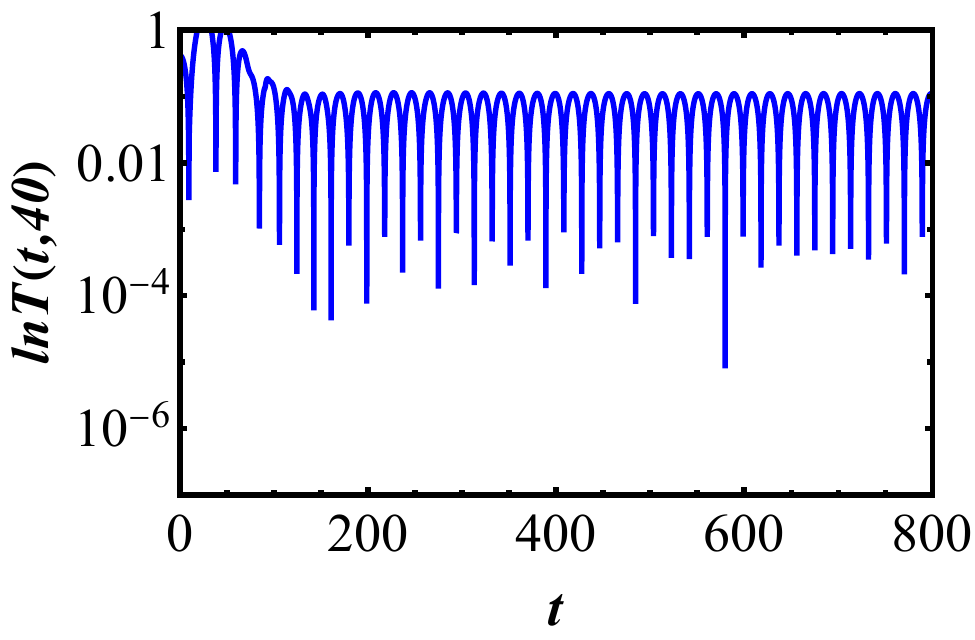} 
			\caption{$T(t,40) = |\psi (t,40)|,n=1$}
			\label{6(c)}
		\end{subfigure}
		\vspace{0.5cm} 
		\begin{subfigure}{0.45\textwidth} 
			\centering
			\includegraphics[width=\linewidth]{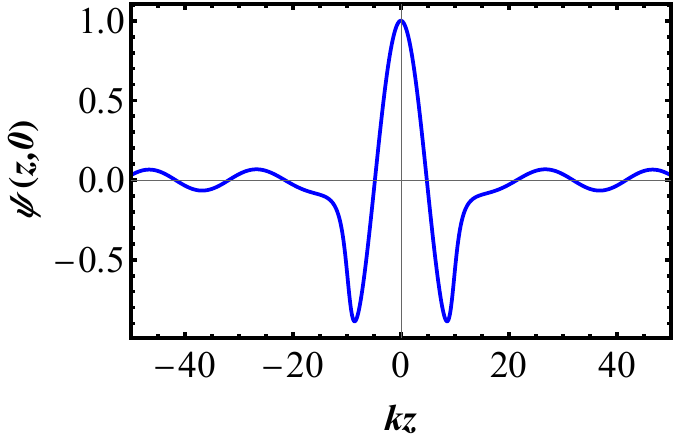} 
			\caption{Initial wave profile, $n=2$}
			\label{6(d)}
		\end{subfigure}
		\hspace{0.025\textwidth} 
		\begin{subfigure}{0.45\textwidth} 
			\centering
			\includegraphics[width=\linewidth]{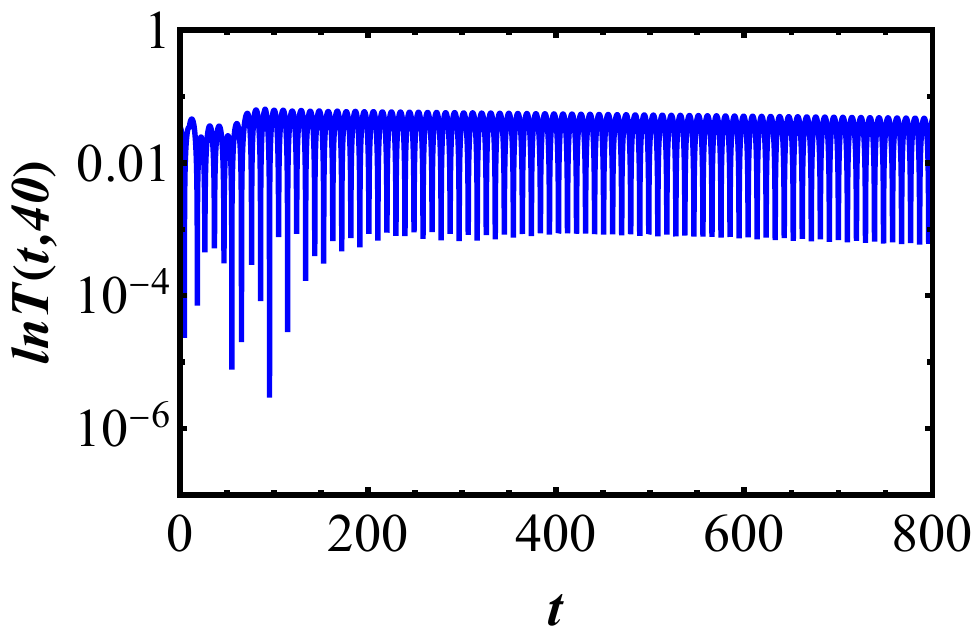} 
			\caption{$T(t,40) = |\psi (t,40)|,n=2$}
			\label{6(e)}
		\end{subfigure}
		\vspace{0.5cm} 
		\begin{subfigure}{0.45\textwidth} 
			\centering
			\includegraphics[width=\linewidth]{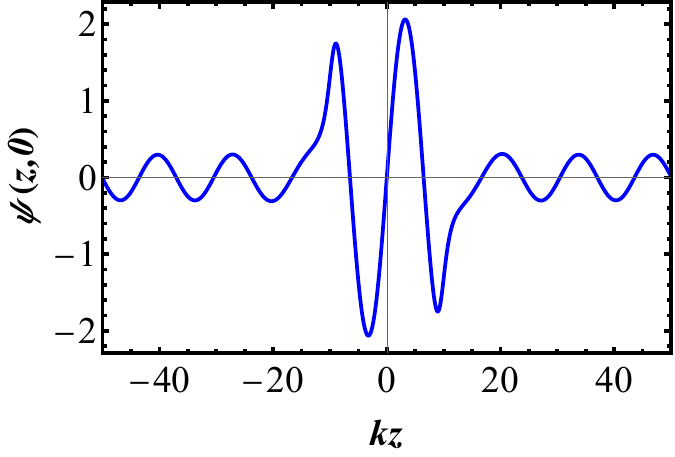} 
			\caption{Initial wave profile, $n=3$}
			\label{6(f)}
		\end{subfigure}
		\hspace{0.025\textwidth} 
		\begin{subfigure}{0.45\textwidth} 
			\centering
			\includegraphics[width=\linewidth]{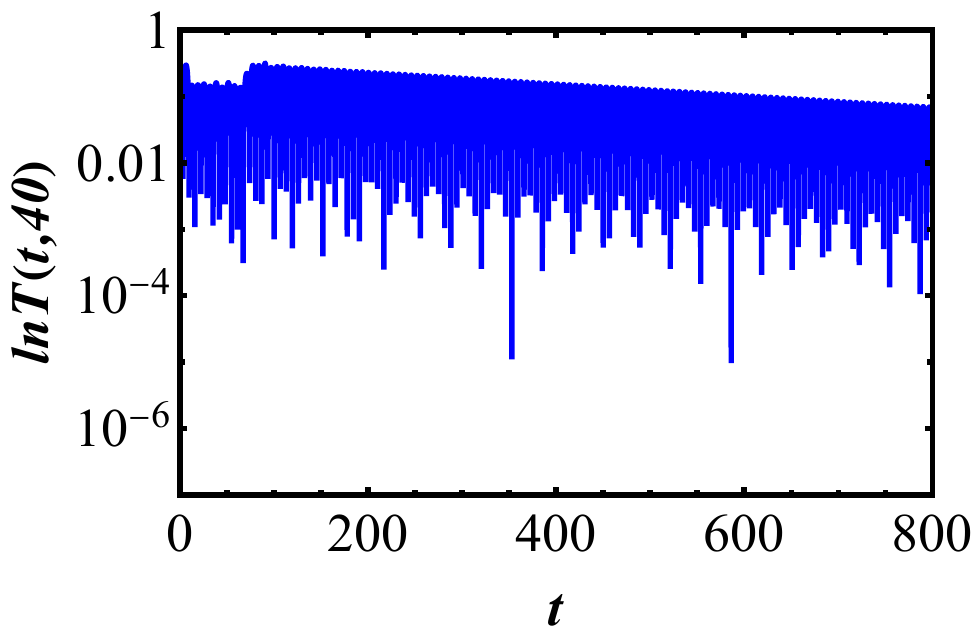} 
			\caption{$T(t,40) = |\psi (t,40)|,n=3$}
			\label{6(g)}
		\end{subfigure}
        
        \vspace{0.5cm} 
		\begin{subfigure}{0.45\textwidth} 
			\centering
			\includegraphics[width=\linewidth]{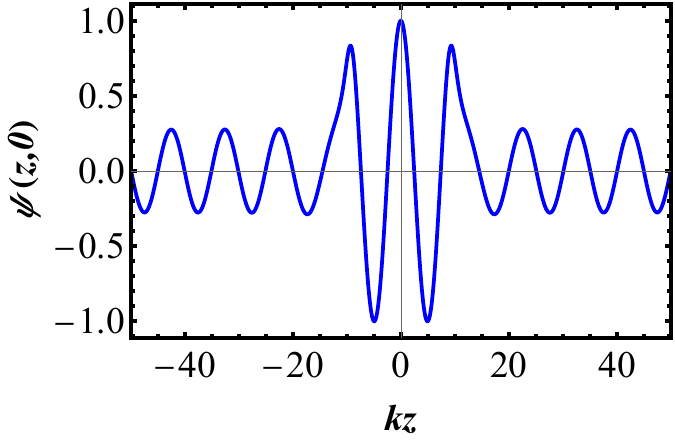} 
			\caption{Initial wave profile, $n=4$}
			\label{6(h)}
		\end{subfigure}
		\hspace{0.025\textwidth} 
		\begin{subfigure}{0.45\textwidth} 
			\centering
			\includegraphics[width=\linewidth]{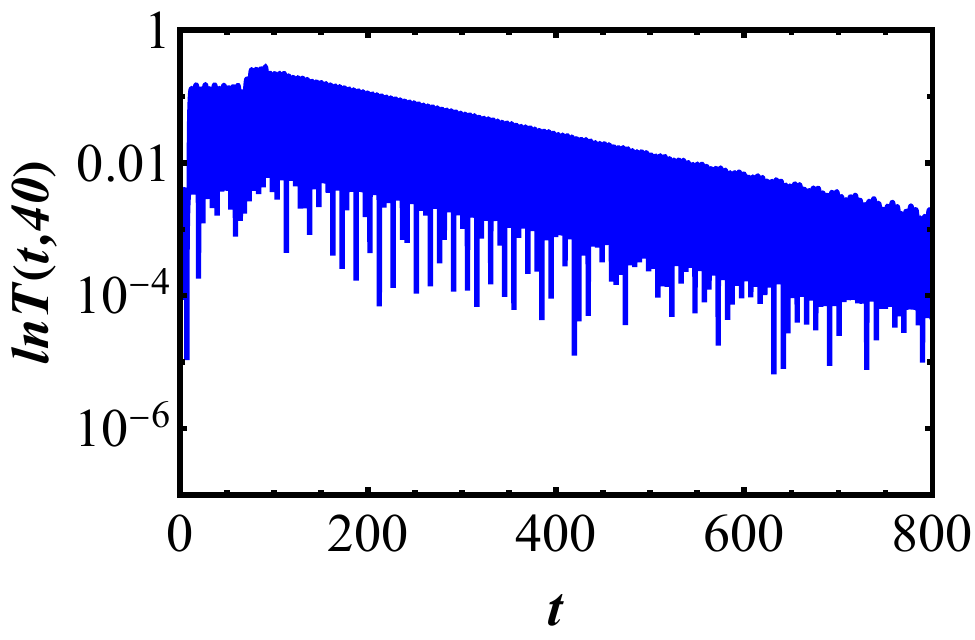} 
			\caption{$T(t,40) = |\psi (t,40)|,n=4$}
			\label{6(i)}
		\end{subfigure}
		\caption{The left column shows the initial wave profile, while the right column presents the logarithmic time evolution of the waveform.}
		\label{Ab10te}
	\end{figure}

Figure~\ref{对比图} shows the transmission spectra of the effective potential for $b=5$ and $b=10$. The transmission peaks are approximately evenly spaced, and the spacing varies with the parameter $b$. We label the resonant modes in order of increasing frequency, starting with $n=1$ for the fundamental mode; the overtone number is therefore $n-1$. The wave profiles corresponding to these transmission peaks, together with their time evolution when used as initial conditions, are shown in Fig.~\ref{波形演化} and Fig.~\ref{Ab10te}. A clear interval of exponential decay is visible. Fitting the waveform over this interval allows us to extract the real and imaginary parts of the QNFs.

To verify the accuracy of our results, we compare them with those obtained from the direct integration method, as shown in Tab.~\ref{tab-I}. The real parts obtained from the numerical evolution and direct integration methods agree to better than $0.06\%$. For most modes, the relative deviation of the imaginary parts is below $2\%$. The exception is the fundamental mode with $b=10$, whose extremely small damping rate leads to a comparatively large relative error in the extracted imaginary part. 

We find that the real parts of the QNFs increase approximately linearly with the resonance order $n$, indicating that the oscillation frequencies of higher resonant modes are larger. Meanwhile, the absolute values of the imaginary parts increase rapidly with $n$, implying that higher resonant modes decay faster and therefore possess shorter lifetimes.

For larger values of the parameter $b$, the real parts of the QNFs become smaller, while the absolute values of the imaginary parts are significantly suppressed. This indicates that increasing $b$ lowers the oscillation frequencies and increases the lifetimes of the resonant modes. Such behavior is consistent with the fact that the width of the double-barrier potential increases with $b$, leading to stronger confinement of the gravitational KK modes.
\begin{table}[htbp]
    \centering
    \renewcommand{\arraystretch}{1.5} %
     \resizebox{0.9\textwidth}{!}{
    \begin{tabular}{|c|c|c|c|}
        \hline
        b & n  & Numerical evolution  & Direct integration  \\ 
        \hline
        \multirow{2}{*}{\centering 5} & 1 & $Re =0.340824 \quad Im = -0.000935$ & $Re = 0.340813 \quad Im = -0.000933$ \\
        & 2 & $Re =0.643071 \quad Im = -0.014171$  & $Re =0.642778 \quad Im = - 0.014176 $ \\
        \hline
        \multirow{4}{*}{\centering 10} & 1 & $Re =0.164899 \quad Im = -0.000154$ &  $Re =0.164807 \quad Im = - 0.000032 $ \\
        & 2 & $Re =0.327195 \quad Im = -0.000385$  & $Re =0.327321 \quad Im = - 0.000391$ \\
        & 3 & $Re =0.484797 \quad Im = -0.002017$  & $Re =0.484746  \quad Im = -0.002030$ \\
        & 4 & $Re =0.636418 \quad Im = -0.007066$   & $Re = 0.636277 \quad Im = - 0.007115$  \\
        \hline
    \end{tabular}%
      }
    \caption{Comparison of the QNFs obtained from numerical evolution and direct integration with $ b = 5 $ and $ b = 10 $ in Solution I.}
    \label{tab-I}
\end{table}

\subsection{Solution II}
In this background solution, we first obtain the kink-like configuration of the scalar field:
\begin{equation}
\phi(y) = \tanh(ky + b) + \tanh(ky - b).
\end{equation}
We then obtain analytical solutions for the warp factor $A(y)$ and the scalar potential $V(y)$. However, since the expression for $A(y)$ contains more than one hundred terms, we show its behavior graphically in this work.

\begin{figure}[htbp]
    \centering
    \includegraphics[width=0.5\textwidth]{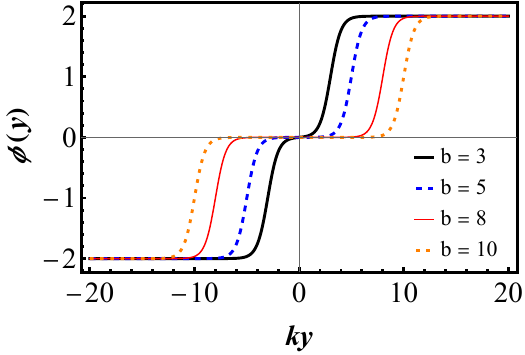}
    \caption{Scalar field in Solution II for different values of the parameter $b$.}
    \label{fig:phi_modelb}
\end{figure}

\begin{figure}[htbp]
		\centering
		\begin{subfigure}{0.45\textwidth} 
			\centering
			\includegraphics[width=\linewidth]{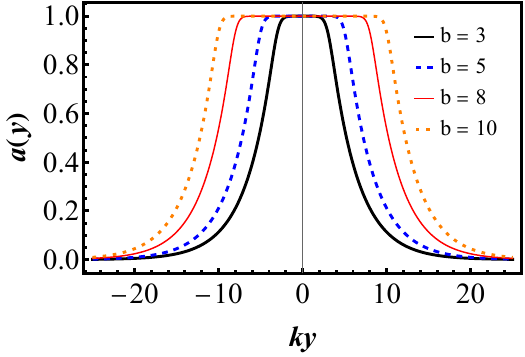} 
			\caption{$\alpha=-0.2$}
			\label{8(b)}
		\end{subfigure}
		\hspace{0.025\textwidth} 
		\begin{subfigure}{0.45\textwidth} 
			\centering
			\includegraphics[width=\linewidth]{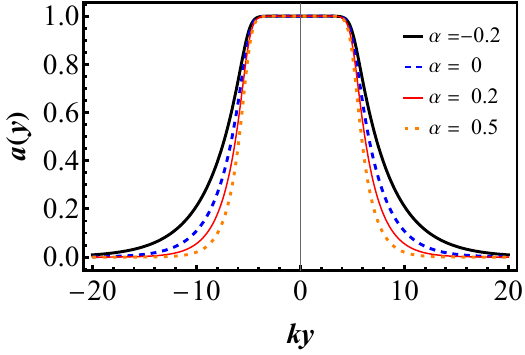} 
			\caption{$b=3$}
			\label{8(c)}
		\end{subfigure}
		\vspace{0.5cm} 
		\begin{subfigure}{0.45\textwidth} 
			\centering
			\includegraphics[width=\linewidth]{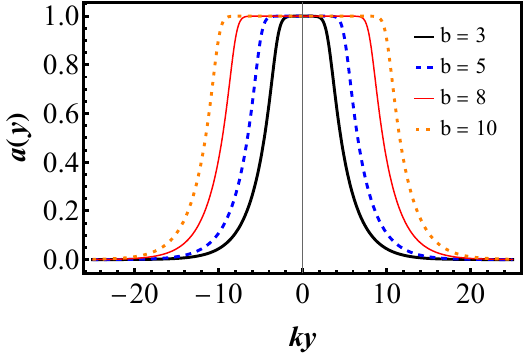} 
			\caption{$\alpha=0$}
			\label{8(d)}
		\end{subfigure}
		\hspace{0.025\textwidth} 
		\begin{subfigure}{0.45\textwidth} 
			\centering
			\includegraphics[width=\linewidth]{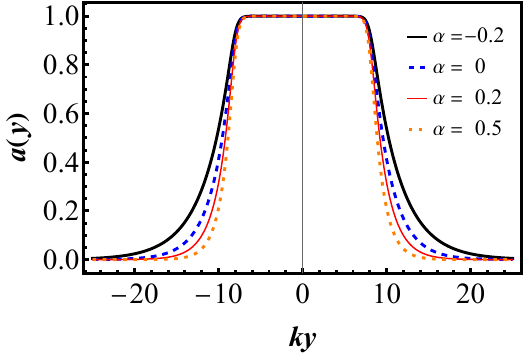} 
			\caption{$b=5$}
			\label{8(e)}
		\end{subfigure}
		\vspace{0.5cm} 
		\begin{subfigure}{0.45\textwidth} 
			\centering
			\includegraphics[width=\linewidth]{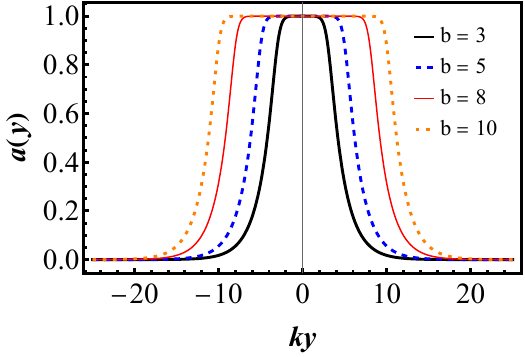} 
			\caption{$\alpha=0.2$}
			\label{8(f)}
		\end{subfigure}
		\hspace{0.025\textwidth} 
		\begin{subfigure}{0.45\textwidth} 
			\centering
			\includegraphics[width=\linewidth]{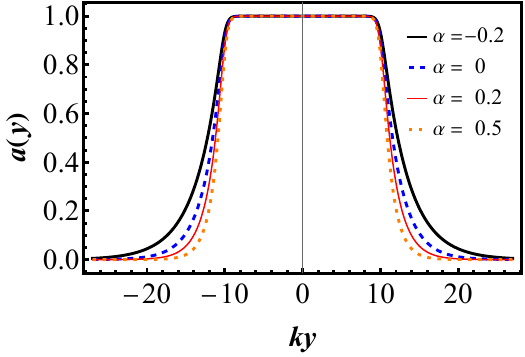} 
			\caption{$b=8$}
			\label{8(g)}
		\end{subfigure}
		\caption{The warp factor $a(y)$ in Solution II with different values of the parameters $b$ and $\alpha$.
		}
		\label{modelbwarp}
	\end{figure}

The profile of the scalar field is shown in Fig.~\ref{fig:phi_modelb}. As in Solution I, the parameter $b$ characterizes the width of the kink-like plateau. However, the warp factor also depends on $\alpha$, as shown in Fig.~\ref{modelbwarp}. The parameter $b$ mainly controls the width of the plateau in the warp factor: increasing $b$ broadens the flat region. In contrast, the parameter $\alpha$ modifies the asymptotic behavior: larger values of $\alpha$ enhance the sharpness of the transition, making the warp factor progressively steeper.

\begin{figure}[htbp]
		\centering
		\begin{subfigure}{0.45\textwidth} 
			\centering
			\includegraphics[width=\linewidth]{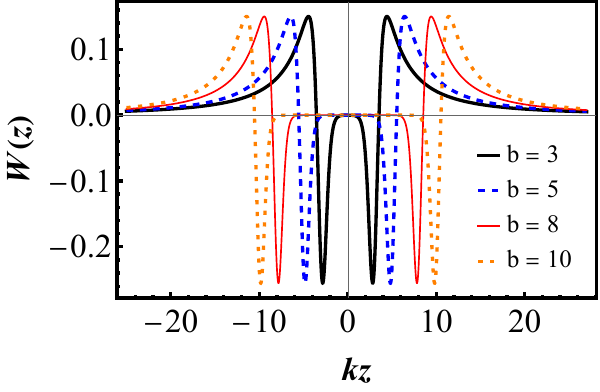} 
			\caption{$\alpha=-0.2$}
			\label{9(b)}
		\end{subfigure}
		\hspace{0.025\textwidth} 
		\begin{subfigure}{0.45\textwidth} 
			\centering
			\includegraphics[width=\linewidth]{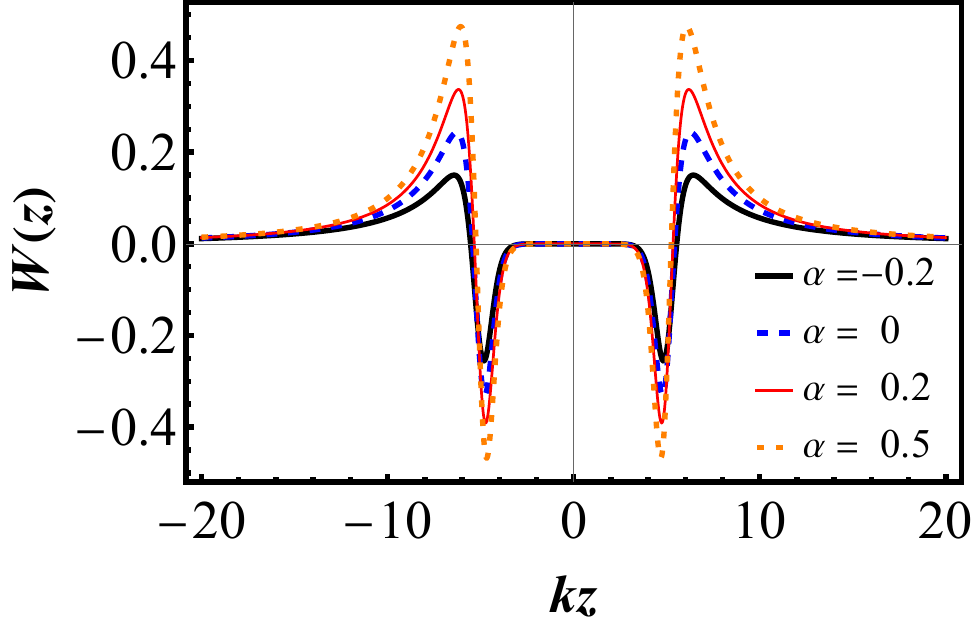} 
			\caption{$b=3$}
			\label{9(c)}
		\end{subfigure}
		\vspace{0.5cm} 
		\begin{subfigure}{0.45\textwidth} 
			\centering
			\includegraphics[width=\linewidth]{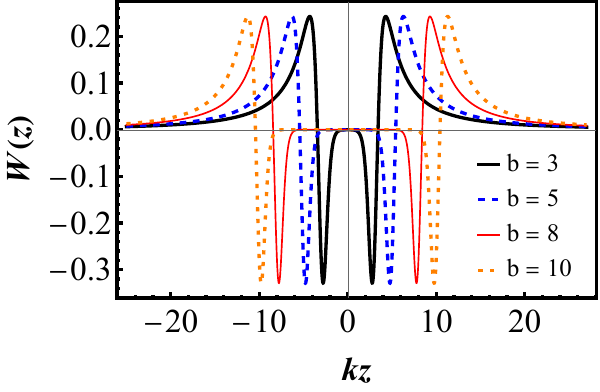} 
			\caption{$\alpha=0$}
			\label{9(d)}
		\end{subfigure}
		\hspace{0.025\textwidth} 
		\begin{subfigure}{0.45\textwidth} 
			\centering
			\includegraphics[width=\linewidth]{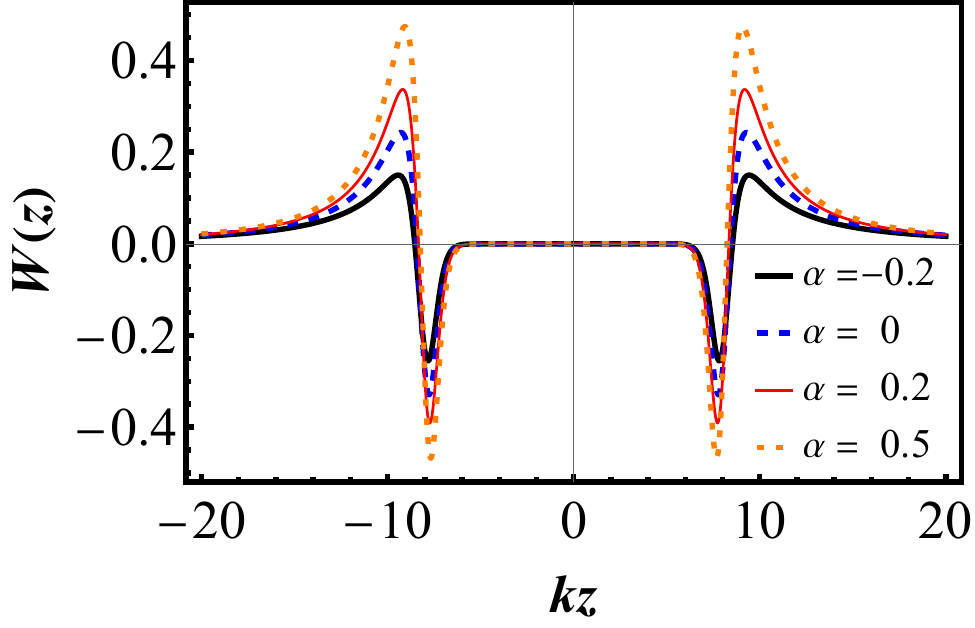} 
			\caption{$b=5$}
			\label{9(e)}
		\end{subfigure}
		\vspace{0.5cm} 
		\begin{subfigure}{0.45\textwidth} 
			\centering
			\includegraphics[width=\linewidth]{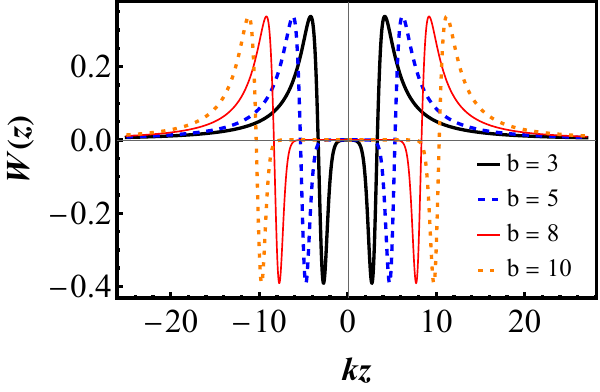} 
			\caption{$\alpha=0.2$}
			\label{9(f)}
		\end{subfigure}
		\hspace{0.025\textwidth} 
		\begin{subfigure}{0.45\textwidth} 
			\centering
			\includegraphics[width=\linewidth]{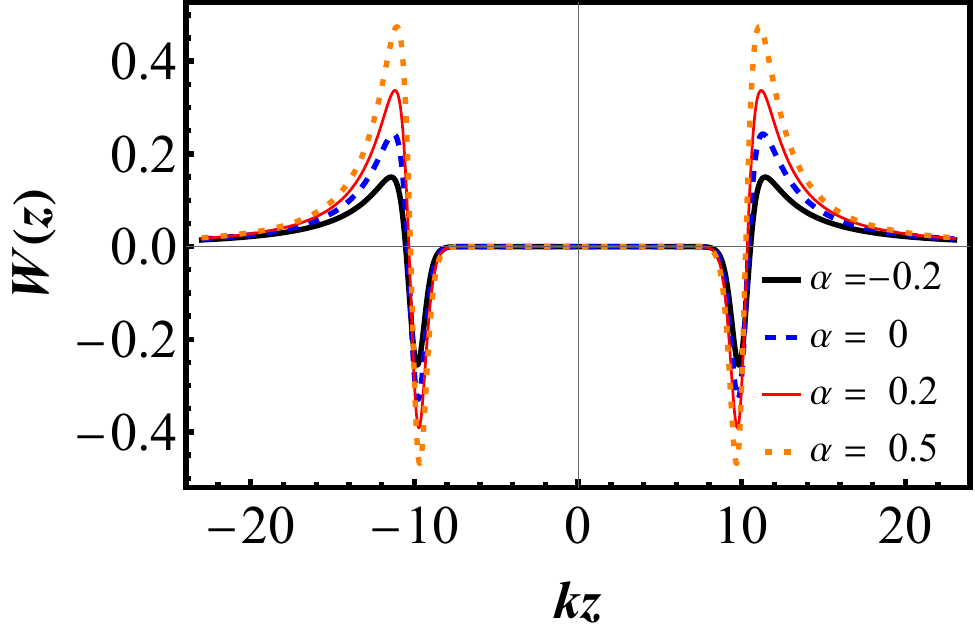} 
			\caption{$b=10$}
			\label{9(g)}
		\end{subfigure}
		\caption{The effective potential $W(z)$ in Solution II with different values of the parameters $b$ and $\alpha$.
		}
		\label{BV}
	\end{figure}

Based on the warp factor solution, we obtain the corresponding effective potential, as shown in Fig.~\ref{BV}. The effective potential exhibits a typical double-barrier structure. The separation between the two barriers increases with the parameter $b$, while the barrier height increases as $\alpha$ becomes larger.

\begin{figure}[htbp]
    \centering
    \hspace{0.0125\textwidth}
    \begin{subfigure}[b]{0.45\textwidth}
        \centering
        \includegraphics[width=\textwidth]{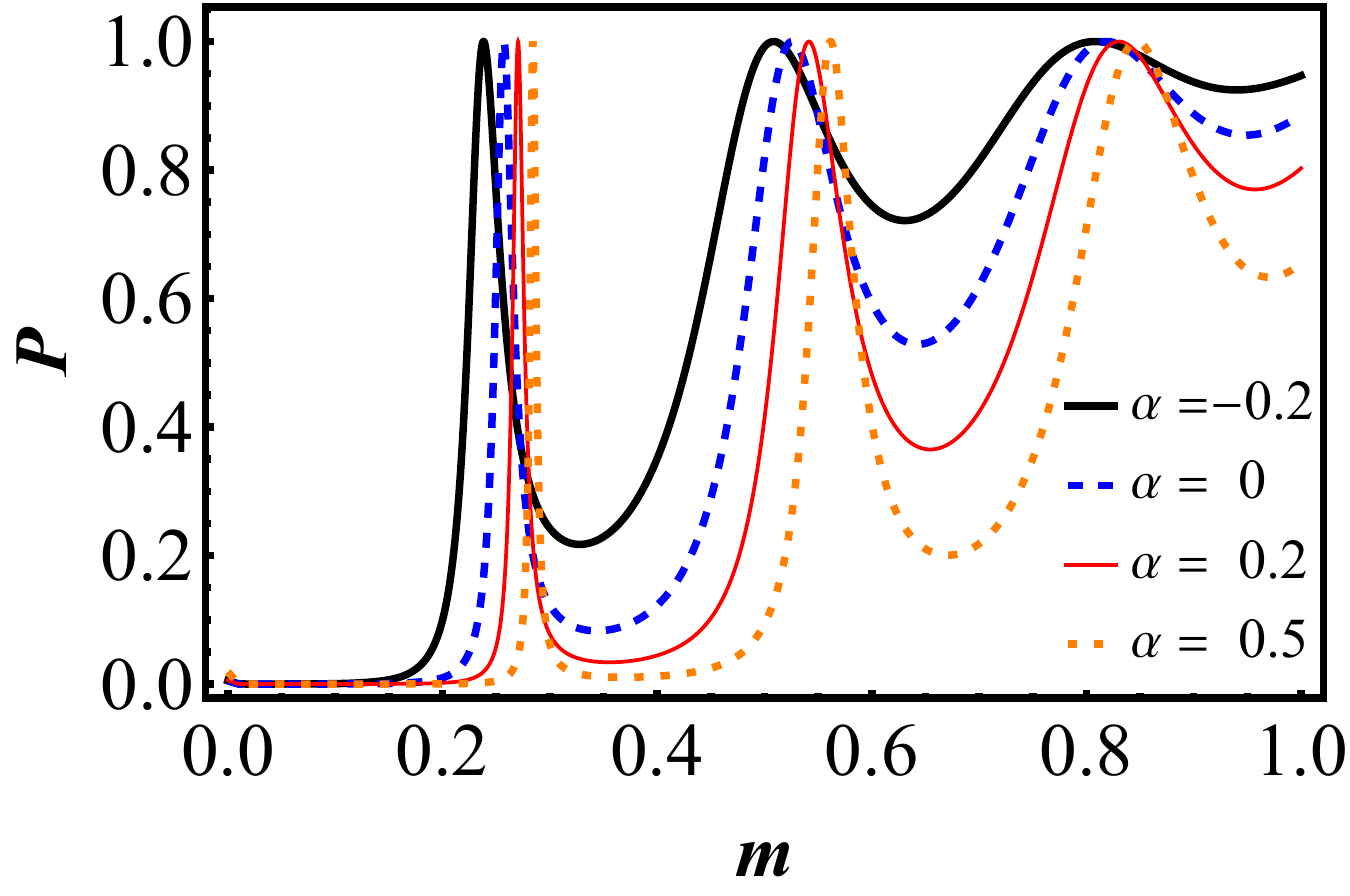} 
            \caption{$b=5$}
    \end{subfigure}
        \hspace{0.0125\textwidth} 
    \begin{subfigure}[b]{0.45\textwidth}
        \centering
        \includegraphics[width=\textwidth]{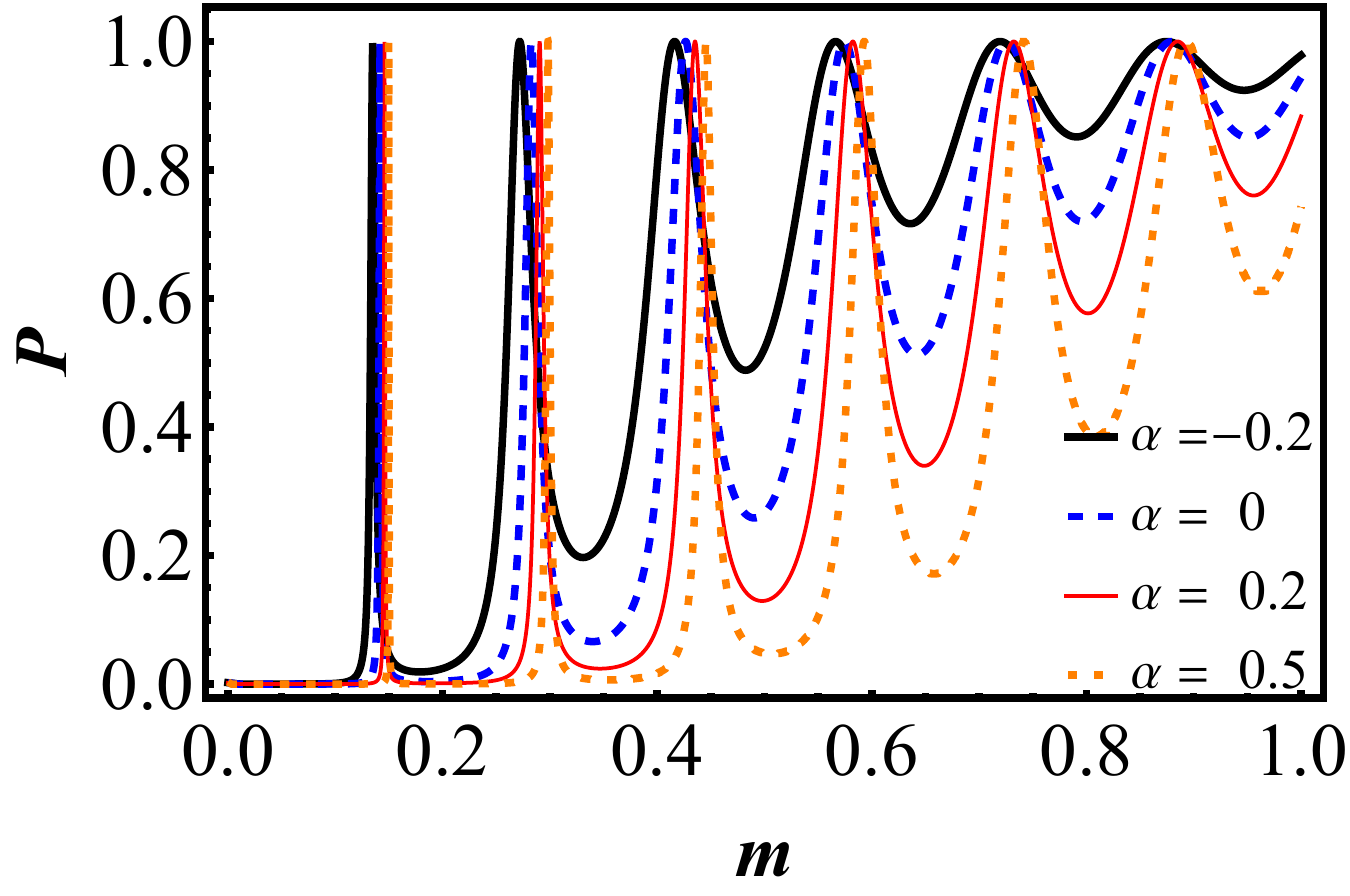} 
        \caption{$b=10$}
    \end{subfigure}
    \caption{The transmission spectrum of the effective potential in Solution II.}
    \label{B对比图}  
\end{figure}
The transmission spectra of the effective potential for different values of $\alpha$ are shown in Fig.~\ref{B对比图}. The peak positions of the transmission spectrum, corresponding to the mass parameter $m$, increase slightly with $\alpha$. This indicates that, for fixed $b$, the parameter $\alpha$ has a slight effect on the width of the resonant cavity. As $\alpha$ increases, the cavity becomes narrower, leading to a shift of the transmission peaks toward larger spectral values. In addition, increasing $b$ leads to a denser distribution of transmission peaks.

\begin{figure}[htbp]
		\centering
		\begin{subfigure}{0.44\textwidth} 
			\centering
			\includegraphics[width=\linewidth]{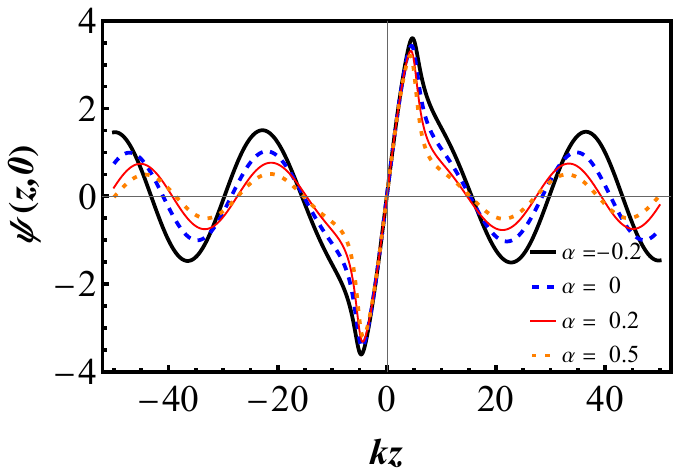} 
			\caption{Initial wave profile, $n=1$}
			\label{10(b)}
		\end{subfigure}
		\hspace{0.025\textwidth} 
		\begin{subfigure}{0.46\textwidth} 
			\centering
			\includegraphics[width=\linewidth]{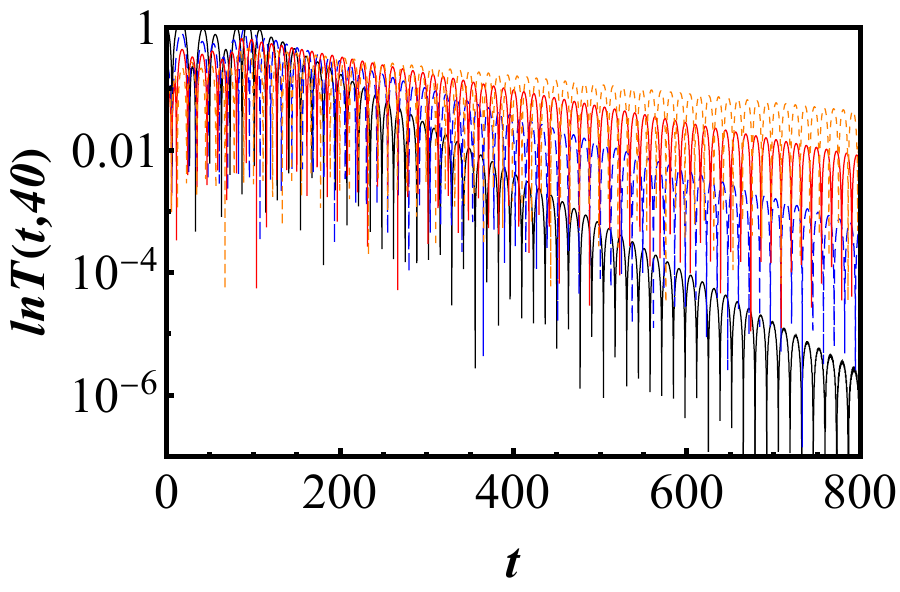} 
			\caption{$T(t,40) = |\psi (t,40)|,n=1$}
			\label{10(c)}
		\end{subfigure}
		\vspace{0.5cm} 
		\begin{subfigure}{0.45\textwidth} 
			\centering
			\includegraphics[width=\linewidth]{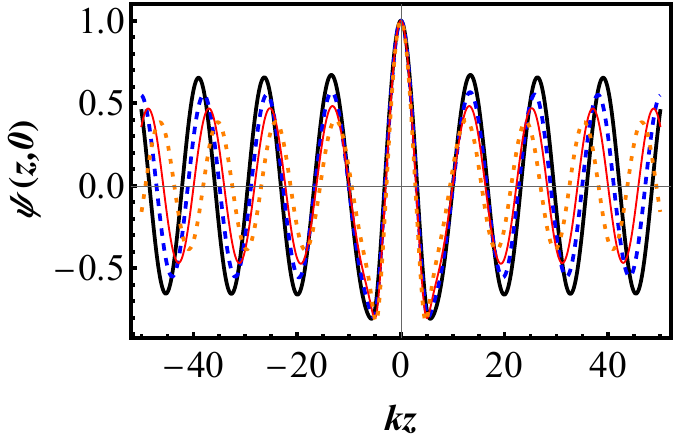} 
			\caption{Initial wave profile, $n=2$}
			\label{10(d)}
		\end{subfigure}
		\hspace{0.025\textwidth} 
		\begin{subfigure}{0.45\textwidth} 
			\centering
			\includegraphics[width=\linewidth]{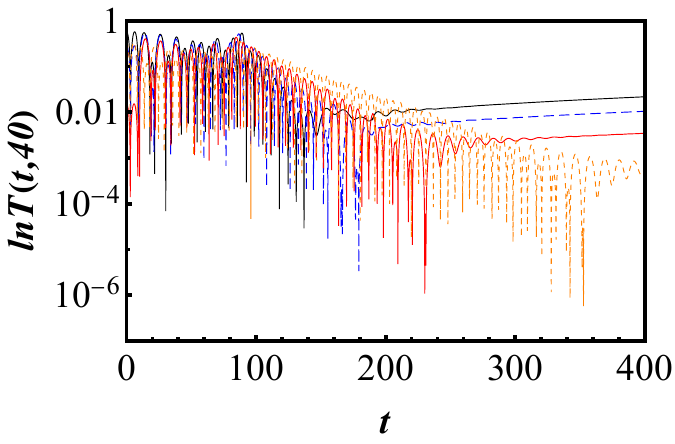} 
			\caption{$T(t,40) = |\psi (t,40)|,n=2$}
			\label{10(e)}
		\end{subfigure}
		\caption{The left column shows the initial wave profile, while the right column presents the logarithmic time evolution of the waveform for $b=5$ with different values of $\alpha$.}
		\label{波形演化B1}
	\end{figure}

Following the same procedure, we obtain the initial wave profiles and the logarithmic time evolution of the waveforms at $kz=40$ for the first two resonant modes ($n=1,2$) with $b=5$ and different values of $\alpha$. As shown in Fig.~\ref{波形演化B1}, for $n=1$, the waveform decays more slowly as $\alpha$ increases. In contrast, the $n=2$ mode exhibits a much faster decay than the $n=1$ mode. The corresponding results are summarized in Tab.~\ref{tab-II}. 
\begin{table}[htbp]
    \centering
    \renewcommand{\arraystretch}{1.5}
    \resizebox{0.9\textwidth}{!}{%
    \begin{tabular}{|c|c|c|c|c|}
        \hline
        $b$ & $\alpha$ & $n$ & Numerical evolution & Direct integration \\
        \hline

        \multirow{8}{*}{\centering 5}
        & \multirow{2}{*}{\centering $-0.2$}
        & 1 & $Re =0.233765 \quad Im = -0.018693$ & $Re =0.234906  \quad Im = -0.018224$\\
        & & 2 & $Re =0.483322 \quad Im = -0.083903$ & $Re =0.489312  \quad Im = -0.079667$\\
        \cline{2-5}

        & \multirow{2}{*}{\centering $0$}
        & 1 & $Re =0.256369 \quad Im = -0.010328$ & $Re =0.255937  \quad Im = -0.010337$\\
        & & 2 & $Re =0.515015 \quad Im = -0.062297$ & $Re =0.515013  \quad Im = -0.053686$\\
        \cline{2-5}

        & \multirow{2}{*}{\centering $0.2$}
        & 1 & $Re =0.270092 \quad Im = -0.006200$ & $Re =0.269896 \quad Im = -0.006406$\\
        & & 2 & $Re =0.534739 \quad Im = -0.038917$ & $Re =0.535271  \quad Im = -0.038267$\\
        \cline{2-5}

        & \multirow{2}{*}{\centering $0.5$}
        & 1 & $Re =0.284142 \quad Im = -0.003378$ & $Re =0.284104  \quad Im = -0.003516$\\
        & & 2 & $Re =0.557267 \quad Im = -0.024663$ & $Re =0.558673  \quad Im = -0.024798$\\
        \hline

        \multirow{8}{*}{\centering 10}
        & \multirow{2}{*}{\centering $-0.2$}
        & 1 & $Re =0.135084 \quad Im = -0.002479$ &  $Re =0.134335 \quad Im =- 0.002399$\\
        & & 2 & $Re =0.269665 \quad Im = -0.013919$ & $Re = 0.271243\quad Im=-0.014471$\\
        \cline{2-5}

        & \multirow{2}{*}{\centering $0$}
        & 1 & $Re =0.141874 \quad Im = -0.001089$ & $Re =0.141503 \quad Im = -0.001049$\\
        & & 2 & $Re =0.282013 \quad Im = -0.007242$ & $Re =0.283016 \quad Im = -0.007604$\\
        \cline{2-5}

        & \multirow{2}{*}{\centering $0.2$}
        & 1 & $Re =0.145966 \quad Im = -0.000567$ & $Re =0.145951 \quad Im = -0.000539$\\
        & & 2 & $Re =0.290696 \quad Im = -0.004136$ & $Re =0.290939  \quad Im = -0.004419$\\
        \cline{2-5}

        & \multirow{2}{*}{\centering $0.5$}
        & 1 & $Re =0.149984 \quad Im = -0.000271$ & $Re =0.150367  \quad Im = -0.000242$\\
        & & 2 & $Re =0.298543 \quad Im = -0.002144$ & $Re =0.298838  \quad Im = -0.002248$\\
        \hline

    \end{tabular}%
    }
    \caption{The QNFs calculated by the numerical evolution method and the direct integration method for different values of $\alpha$ in Solution II.}
    \label{tab-II}
\end{table}

Due to numerical errors and other effects, the initial wave does not evolve at a single frequency. As the mode associated with the target frequency decays, the evolution gradually becomes dominated by components with longer lifetimes (smaller decay rates). This effect becomes more significant for higher overtones. Thus, the wave function can be well approximated by a superposition of QNMs:
\begin{equation}
    \Psi(t) \simeq \sum_n A_n e^{-i( \omega_n t+ \theta_n)},
\end{equation}
where the complex frequency $\omega_n$ encodes both oscillation and decay, and $\theta_n$ is the initial phase of the $n$-th mode. Due to the $\mathbb{Z}_2$ symmetry of the effective potential, the spatial operator $-\partial_z^2 + W(z)$ commutes with the parity operator. Therefore, parity is a conserved quantity, and the perturbation space can be decomposed into two decoupled subspaces corresponding to even and odd parity, respectively. Thus, we have
\begin{align}
    \Psi_{\text{odd}}(t) &\simeq \sum_{i=1,3,5,\dots} A_i e^{-i\left(\omega_i t + \theta_i\right)},\\
    \Psi_{\text{even}}(t) &\simeq \sum_{j=2,4,6,\dots} A_j e^{-i\left(\omega_j t + \theta_j\right)}.
\end{align}
As shown in Tab.~\ref{tab-I}, the damping rate of the QNMs increases with the resonance order $n$. Therefore, at sufficiently late times, the dominant contamination in extracting the $n$-th QNM arises from lower-order modes with the same parity. Accordingly, in the late-time regime, the waveform evolved from the fourth resonant mode can be approximately expressed as 
\begin{equation}
\Psi_4(t) \simeq
A_{2}e^{-i\left(\omega_{2}t+\theta_{2}\right)}+A_{4}e^{-i\left(\omega_{4}t+\theta_{4}\right)}.\label{fitttt}
\end{equation}
\begin{figure}[htbp]
		\centering
		\begin{subfigure}{0.44\textwidth} 
			\centering
			\includegraphics[width=\linewidth]{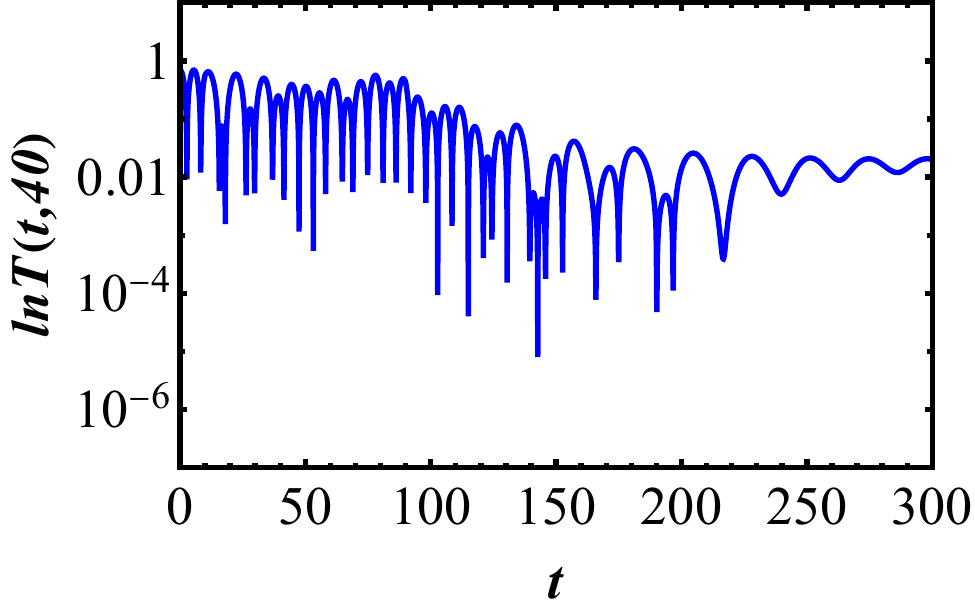} 
			 \caption{$b=10,n=4,\alpha=-0.2$}
			\label{11(b)}
		\end{subfigure}
		\hspace{0.025\textwidth} 
		\begin{subfigure}{0.46\textwidth} 
			\centering
			\includegraphics[width=\linewidth]{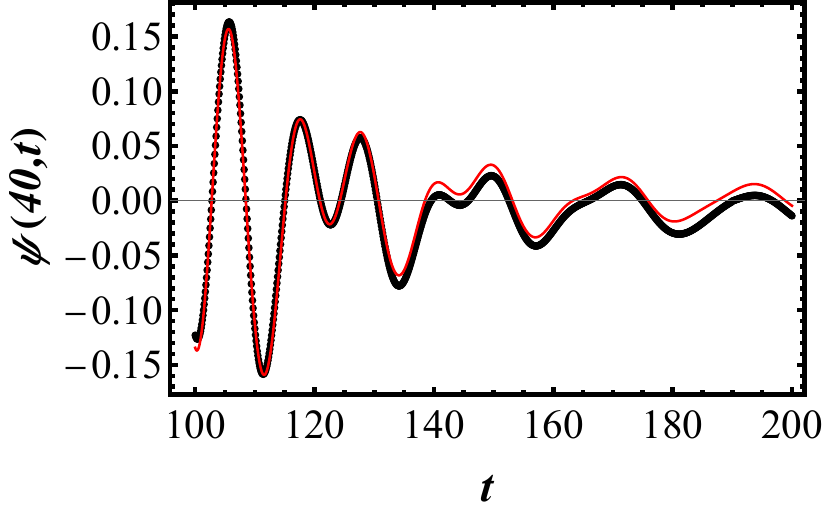} 
			 \caption{$b=10,n=4,\alpha=-0.2$}
			\label{11(c)}
		\end{subfigure}
		\vspace{0.5cm} 
		\begin{subfigure}{0.45\textwidth} 
			\centering
			\includegraphics[width=\linewidth]{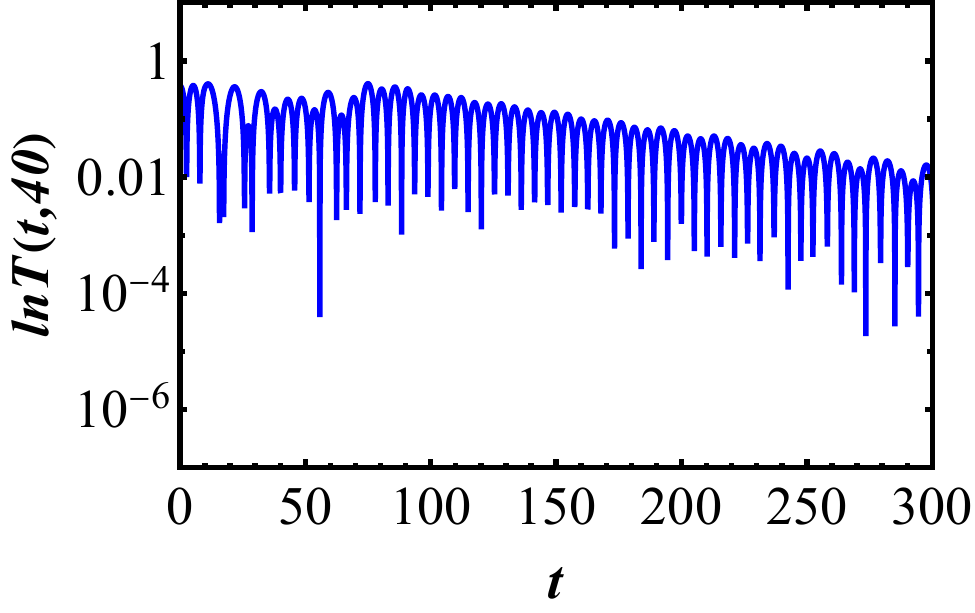} 
			 \caption{$b=10,n=4,\alpha=0.5$}
			\label{11(d)}
		\end{subfigure}
		\hspace{0.025\textwidth} 
		\begin{subfigure}{0.45\textwidth} 
			\centering
			\includegraphics[width=\linewidth]{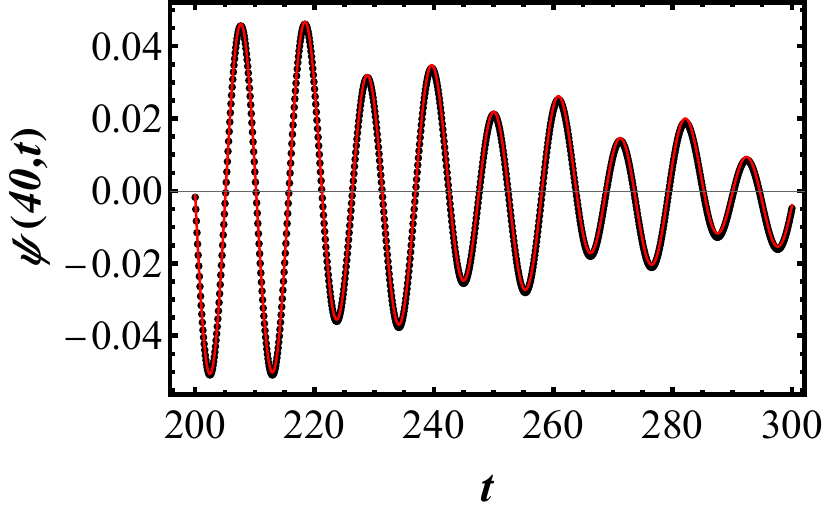} 
			 \caption{$b=10,n=4,\alpha=0.5$}
			\label{11(e)}
		\end{subfigure}
		\caption{The left column shows the logarithmic representation of the time-evolved waveform, while the right column presents a comparison between the time evolution of the waveform (black curve) and the numerical fitting result (red curve).
		}
		 \label{a对比图}  
	\end{figure}
By fitting the time-dependent waveform with Eq.~\eqref{fitttt}, we can extract the oscillation frequency and damping rate of the QNM with $n=4$. The same fit also yields the oscillation frequency and damping rate of the QNM with $n=2$.
The fitting results are shown in Fig.~\ref{a对比图}, while additional results are presented in Tab.~\ref{tab-III}. 
\begin{table}[htbp]
    \centering
    \renewcommand{\arraystretch}{1.5}
    \resizebox{0.9\textwidth}{!}{%
    \begin{tabular}{|c|c|c|c|c|}
        \hline
        $b$ & $\alpha$ & $n$ & Numerical evolution & Direct integration \\
        \hline
        \multirow{4}{*}{\centering 10}
        & \multirow{2}{*}{\centering $-0.2$}
        & 2 & $Re = 0.271637 \quad Im = -0.014972$ &  $Re =0.271243 \quad Im =- 0.014471$\\
        & & 4 & $Re =0.560978 \quad Im = -0.051454$ & $Re = 0.559977\quad Im=-0.049857$\\
        \cline{2-5}
        & \multirow{2}{*}{\centering $0.5$}
        & 2 & $Re =0.297853\quad Im = -0.002144$ & $Re =0.298838  \quad Im = -0.002248$\\
        & & 4 & $Re =0.592224 \quad Im = -0.015549$ & $Re =0.591962   \quad Im = -0.015203$\\
        \hline
    \end{tabular}%
    }
    \caption{The QNFs calculated by the numerical evolution method and the direct integration method for different values of $\alpha$ in Solution II.}
    \label{tab-III}
\end{table}

A natural question is how this method performs for higher overtones. To further assess its applicability, we consider the case $b=10$ and $\alpha=0.5$ as an example and attempt to identify the highest overtones that can be reliably extracted. We fit the waveforms using the following ansatz:
\begin{equation}
\begin{aligned}
\Psi_1(t) &\simeq A_{1} e^{-i(\omega_{1} t + \theta_{1})}, \\
\Psi_2(t) &\simeq A_{2} e^{-i(\omega_{2} t + \theta_{2})}, \\
\Psi_3(t) &\simeq A_{1}e^{-i(\omega_{1}t+\theta_{1})}+A_{3}e^{-i(\omega_{3}t+\theta_{3})},\\
\Psi_4(t) &\simeq A_{2}e^{-i(\omega_{2}t+\theta_{2})}+A_{4}e^{-i(\omega_{4}t+\theta_{4})},\\
\Psi_5(t) &\simeq A_{1}e^{-i(\omega_{1}t+\theta_{1})}+A_{3}e^{-i(\omega_{3}t+\theta_{3})}+A_{5}e^{-i(\omega_{5}t+\theta_{5})},\\
\Psi_6(t) &\simeq A_{2}e^{-i(\omega_{2}t+\theta_{2})}+A_{4}e^{-i(\omega_{4}t+\theta_{4})}+A_{6}e^{-i(\omega_{6}t+\theta_{6})}.
\end{aligned}
\end{equation}

For higher overtones ($n \geq 7$), the extraction procedure becomes increasingly challenging. This difficulty is mainly due to the significantly larger damping rates, which result in a very short-lived oscillatory regime. Consequently, the waveform quickly becomes dominated by non-oscillatory contributions such as numerical noise and late-time tails, making the identification of the target mode unreliable.

To mitigate this issue, an additional phenomenological background term is introduced in the fitting ansatz to model the non-oscillatory components and improve the isolation of higher QNM contributions. We illustrate this procedure by considering the mode with $n=7$:
\begin{equation}
\Psi_7(t) \simeq A_0 e^{-w_0 t}
+ A_{1}e^{-i(\omega_{1}t+\theta_{1})}
+ A_{3}e^{-i(\omega_{3}t+\theta_{3})}
+ A_{5}e^{-i(\omega_{5}t+\theta_{5})}
+ A_{7}e^{-i(\omega_{7}t+\theta_{7})}.
\end{equation}

Due to the presence of background decay terms and numerical errors, the raw waveform is significantly contaminated in the high-overtone regime. To further reduce these effects, additional phenomenological background terms are included in the fitting ansatz to capture non-oscillatory contributions. Within this improved framework, we are able to extract the mode with $n=8$. The corresponding fitting results are presented in Tab.~\ref{tab:IV}. As discussed above, the results for $n=8$ deviate more significantly from those obtained by direct integration than do the results for lower overtones.
\begin{table}[htbp]
    \centering
    \renewcommand{\arraystretch}{1.5}
    \resizebox{0.9\textwidth}{!}{
    \begin{tabular}{|c|c|c|}
        \hline
        $n$ & Numerical evolution & Direct integration \\
        \hline
        1 & $Re =0.149984 \quad Im = -0.000271$ & $Re =0.150367 \quad Im = -0.000242$\\
        2 & $Re =0.298543 \quad Im = -0.002144$ & $Re =0.298838 \quad Im =-0.002248$\\ 
        3 & $Re =0.445372 \quad Im = -0.006885$ & $Re =0.443831 \quad Im = -0.007005$\\
        4 & $Re =0.592224 \quad Im = -0.015549$ & $Re =0.591962 \quad Im = -0.015203$\\
        5 & $Re =0.739781 \quad Im = -0.027913$ & $Re =0.739971 \quad Im = -0.027165$\\
        6 & $Re =0.887491 \quad Im = -0.044125$ & $Re =0.889016 \quad Im =-0.042289$\\
        7 & $Re =1.045446 \quad Im = -0.063525$ & $Re =1.039672 \quad Im =-0.059524$\\
        8 & $Re =1.244312 \quad Im = -0.088599$ & $Re =1.192054 \quad Im =-0.077877$\\
        \hline
    \end{tabular}
    }
    \caption{Comparison of the QNFs obtained from the numerical evolution and direct integration methods for $b=10$ and $\alpha=0.5$ in Solution II.}
    \label{tab:IV}
\end{table}

\section{Conclusion and Discussion}
	\label{sec:conclusion}
In this work, we have investigated tensor perturbations in a thick-brane scenario within the $f(R,T)=R+\alpha T$ gravity model. Starting from the background solutions of the warp factor and scalar field, we constructed the corresponding effective potential governing tensor perturbations. We find that the effective potential exhibits a typical double-barrier structure, whose properties are strongly controlled by the model parameters $b$ and $\alpha$.

We analyzed the effects of these parameters in detail. The parameter $b$ mainly determines the spatial separation between the two potential barriers, with larger values leading to a wider effective potential. In contrast, the parameter $\alpha$ mainly controls the height of the barriers and modifies the asymptotic behavior of the potential. In particular, increasing $\alpha$ enhances the overall barrier height and shifts the potential profile.

Based on the effective potential, we studied the transmission spectra using the transfer matrix method. We found that the transmission peaks are closely related to the real parts of the QNFs. The peak positions increase with $\alpha$, indicating that higher barrier heights correspond to larger oscillation frequencies. Meanwhile, increasing $b$ leads to a denser distribution of transmission peaks, reflecting the widening of the effective potential.

To extract the QNMs, we performed numerical time evolution. Owing to the $\mathbb{Z}_2$ symmetry of the effective potential, the perturbation spectrum naturally decomposes into even and odd parity sectors, which evolve independently. By fitting the late-time waveform with a truncated QNM expansion, we successfully extracted the oscillation frequencies and damping rates of the low-lying modes.

A comparison between numerical evolution and the direct integration method shows good agreement for low-lying modes. However, the discrepancy becomes more significant for higher overtones. This is mainly due to the rapid increase in damping rates, which leads to a very short-lived oscillatory regime. As a result, late-time evolution is increasingly contaminated by numerical noise and non-oscillatory tail contributions, making higher-mode extraction more challenging.

To further investigate the applicability of the method at higher overtone levels, we considered the case $b=10$ and $\alpha=0.5$. We constructed a systematic fitting scheme in which low and intermediate modes are successively included, taking into account parity selection. For $n \geq 7$, the extraction becomes unstable unless additional phenomenological background terms are introduced to model non-oscillatory contributions. Within this improved fitting framework, we were able to extract the modes with $n=7$ and $n=8$. However, we find that the $n=8$ mode exhibits a noticeably larger deviation from the direct integration results compared with lower modes, indicating the increasing contamination and mode mixing in the high-overtone regime.

Overall, the real parts of the QNFs increase approximately linearly with the mode index $n$, while the magnitudes of the imaginary parts also increase, implying faster decay for higher modes. Moreover, larger values of $b$ tend to suppress the damping rate and lead to longer-lived resonant states, while increasing $\alpha$ generally enhances the oscillation frequency by raising the effective potential barrier.

In conclusion, our results show that the $f(R,T)$ correction plays a nontrivial role in shaping the structure of the effective potential and the corresponding quasinormal spectrum. From a gravitational wave perspective, these effects provide a possible imprint of modified gravity in braneworld scenarios. However, the extraction of higher overtones is limited by numerical contamination and mode mixing, suggesting that more refined methods or alternative time-evolution schemes are required for robust identification of high-$n$ QNMs. Future work may extend this analysis to scalar and vector perturbations, as well as more general $f(R,T)$ models.

	\section*{Acknowledgments}
	  This work was supported by the National Natural Science Foundation of China (Grants No.~12475056, No.~12247101), the 111 Project under (Grant No.~B20063), the Natural Science Foundation of Gansu Province (No.~22JR5RA389, No.~25JRRA799),  the Scientific Research Projects of Jingchu University of Technology (Grant No.YYZ202617) and Gansu Province's Top Leading Talent Support Plan, the Fundamental Research Funds for the Central Universities (Grants No. lzujbky-2025-it05).

	\bibliographystyle{unsrt} 
	\bibliography{ref}

\end{document}